\documentclass[prb,showpacs,showkeys,preprintnumbers,superscriptaddress,a4paper,twocolumn]{revtex4-1}
\usepackage[T1]{fontenc}
\usepackage[utf8]{inputenc}
\usepackage{graphicx}
\usepackage{amsmath,amssymb}
\usepackage{amstext}		
\usepackage{blindtext}
\usepackage{enumitem}
\usepackage{psfrag}
\usepackage{xcolor}
\usepackage{epsfig}
\usepackage[colorlinks,citecolor=blue,
    filecolor=blue,
    linkcolor=blue,
    urlcolor=blue]{hyperref}

   \let\c=\chi

\def\beq{\begin{equation}}
\def\eeq{\end{equation}}
\def\bea{\begin{eqnarray}}
\def\eea{\end{eqnarray}}
\def\ba{\begin{array}}
\def\ea{\end{array}}

\begin{document}
\title{An introduction to  Kitaev model-I}
\author{Saptarshi Mandal }\email{saptarshi@iopb.res.in}
\affiliation{Institute of Physics, P.O.: Sainik School, Bhubaneswar 751005, Odisha, India.}
\affiliation{Homi Bhabha National Institute, Mumbai - 400 094, Maharashtra, India}
\author{Arun M Jayannavar }\email{jayan@iopb.res.in}
\affiliation{Institute of Physics, P.O.: Sainik School, Bhubaneswar 751005, Odisha, India.}
\affiliation{Homi Bhabha National Institute, Mumbai - 400 094, Maharashtra, India}

\begin{abstract}
This  pedagogical article is aimed to the beginning graduate students interested in broad field of frustrated magnetism. 
We introduce and present some of the exact results obtained in Kitaev model.  The Kitaev model embodies an unusual two spin interactions 
yet exactly solvable model in two dimension. This exact solvability renders it to realize many emergent many body phenomena  such as $Z_2$ gauge field, spin liquid states, spin fractionalization, topological order exactly. First  we present 
the exact solution of Kitaev model using Majorana fermionisation and elaborate  in detail the $Z_2$ gauge structure. Following this we discuss exact calculation of magnetization, spin-spin correlation function establishing its spin-liquid character.  Spin fractionalization and de-confinement of Majorana fermion is  explained in detail.  Existence of long range multi-spin correlation function and topological degeneracy are discussed to elucidate the  entangled and topological nature of any  eigenstate. Some elementary questionnaires are provided in appropriate places for assimilation of the technical details.  
\end{abstract}

\date{\today}

\maketitle
\section{Introduction}

With the advent of time, discovaries of various new states of matter\cite{goodstein,wen0} continues to surprise and challenge the knowledge and understanding of mankind.  For long time we were familiar with solid, liquid and gaseous states of matter.  However, gradually we discovered that there are subtle differences in the microscopic organizations within various solid or liquid or gaseous states and this needs further classifications and characterizations. For example, a solid can be a crystal or amorphous, can be metal, insulator or semiconductor. In this article we are going to  understand quantum spin liquid states \cite{Bal,kivelson} which for quite some time puzzled the present condensed matter community for its unique property and possible application for future technology. We would take some effort to  understand the each term of the phrase "quantum spin liquid". In ordinary liquid states like water we are concerned about the positions and orientations of each water molecule with respect to other neighbouring water molecule. In fact it is the
relative position and orientation of each water molecule with respect to all other water molecule that defines a given state like solid, liquid or gas.  In this context some preliminary idea like symmetry, long range order etc are important.  Take the example of water in liquid state where the position of a water molecule is in constant change with respect to any other water molecule though a very short range correlation exist. Thus we say that there is no long range order in liquid states. On the other hand   water in ice states has a long range order. Secondly if we look at microscopically around a given  water molecule, the neighbouring environment or the positions of surrounding water molecules looks identical. However this is completely different
in solid material with crystal structure.  We say that for liquid state, the rotational symmetry is present and for solid with a given crystal structure, the continuous rotational symmetry is broken and becomes discrete.  This idea of long range order and symmetry helps us to classify various states of matter \cite{landau}. Fundamental degrees that is relevant in this article   are spins of molecule, atoms or even electrons. It may happen that the molecules or atom that constitute a particular materials may constitute a perfect crystals or ordered solid phases of matter. In that case what we are interested is the collective states of individual spins of the atoms/molecules.  In other case it may happen that the spins of the valence electron that  are roaming freely inside the materials.  The collective states of such large collection of spins can shows diverse states  having long range order or 
absence of it. Ferromagnetic state or paramagnetic states are example of that respectively. However there is a possibility of states of matter  governed by the fundamental property of quantum mechanics  which makes it possible to have states beyond simple ordered collinear states of paramagnetic states. The superposition principle enables this miracles due to which the system can be in a given time, probabilistically, in many states. For example $|\Psi_i \rangle$ is a many-body spin state for the  system where spins at a given site `$p$' is in a state $|i_p \rangle$.  It is possible to have another many-body spin configuration given by state $| \Psi_j \rangle$ where spins at a given site `$p$' is in a state $|j_p \rangle$. $|\Psi_i \rangle$ is said to be different then $| \Psi_j \rangle$ where $|i_p \rangle \neq |j_p \rangle $ for at least a given site `$p$'. One can easily construct different non-equivalent many-body spin configuration.   The fundamental postulate of quantum mechanics says that it is possible for the system to be in state
$| \Psi \rangle = \sum_i a_i |\Psi_i \rangle$ where $ \sum_i |a_i|^2 =1$ \cite{sakurai}.  The possibility that such construction is possible is at the heart of quantum spin-liquid states.  If an experiment is done on such system to investigate the many-body spin configurations of the collective spins, one finds different outcome at different times but with certain probability.    However all the $| \Psi_i \rangle$ that appears in the expressions of $| \Psi \rangle$ are not arbitrarily chosen. They are governed by the microscopic Hamiltonian or mutual interactions among the spins.  There are specific relations among all the $| \Psi_i \rangle$ that appears in the expressions of $| \Psi \rangle $. Such relations can be of diverse nature and very different from model to model or system to system. This particular aspect is known as entanglement. In the present study, in view of the Kitaev model, we would like to explain some concepts and advanced idea of the modern condensed matter physics and  give examples of other system in comparison with Kitaev model.  We  begin by introducing Kitaev model in section \ref{section-1}. Here we describe in detail the exact solution of Kitaev model, its $Z_2$ gauge structure and ground state wave function explicitely. Various exercizes are also given for beggining graduate students to assimilate technical details. Doing this would help appreciating the Kitaev model further.   In section \ref{sec-order}, we evaluate various order parameter such as magnetization, two-spin correlation function to establish the spin-liquid character of Kitaev model. Multispin correlation function and toplological degeneracy is also explained in a simple way to show the long range entanglement hidden in any eigenstate.  We present a brief recapitulation and discuss some recent development for further reading in section \ref{dis}.\\
 
\section{ Kitaev model}
\label{section-1}
\begin{figure}
\psfrag{a}{$\vec{a}_1$}
\psfrag{b}{$\vec{a}_2$}
\includegraphics[width=0.85\linewidth]{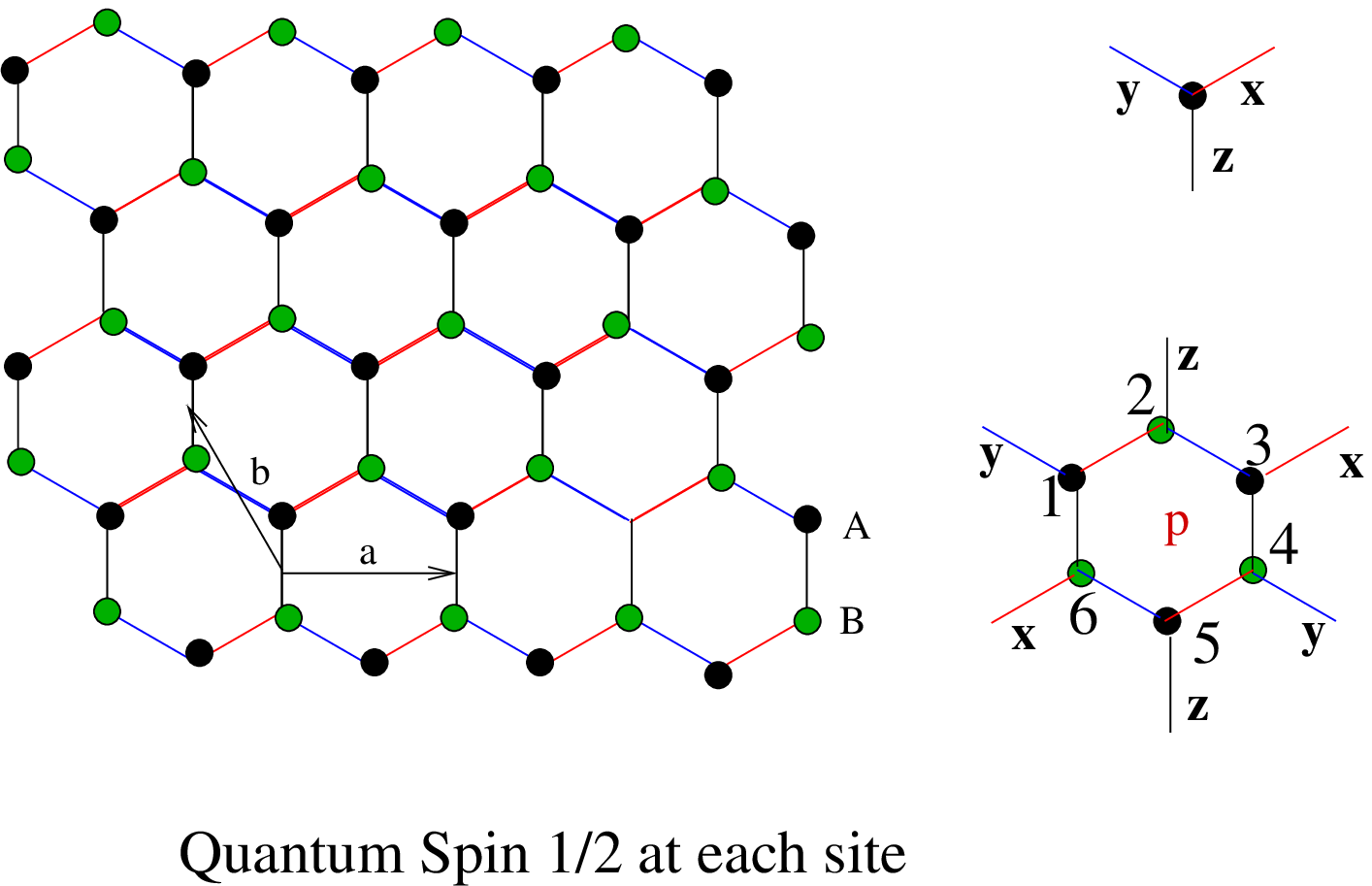}
\caption{\label{fig1} In the left we have shown a cartoon picture of honeycomb lattice. Black and green filled circles shows
two sub lattices that honeycomb lattice is made of. The green, black and red bonds respectively contain
$x$-type, $y$-type and $z$-type interactions respectively. `$\vec{a}_1$' and `$\vec{a}_2$' are two basis vectors. In the lower right corner an elementary hexagonal plaquette is shown with a particular enumeration of sites which has been used to define
the conserved quantity $B_p$ in Eq.~\ref{bpeq}.}
\end{figure}

 At zero temperature, a quantum mechanical system minimizes its energy by occupying its ground states. Hamiltonian is the operator form of energy whose eigenvalues and eigenstates determines all the property of the system in principle \cite{sakurai}. Lets consider a particular Hamiltonian that is of our interest in this article, so called Kitaev Hamiltonian \cite{kitaev-2006}. The system is defined on a honeycomb lattice and at each site of this lattice there is a spin-1/2 particle.  The spin 1/2 particles interact with each other in a specific way.  The origin of such interaction is beyond the scope of this pedagogical article  and interested readers are requested to have a look at the recent progresses \cite{winter}. We will only discuss  few elementary  reasons and consequences of the origin of interactions which involves spin-magnetic moments \cite{mermin}. Lets consider the situation where we bring in two hydrogen atoms close to each other from an initial position where they were at large separation.  Hydrogen atom is made of one proton and one electron.  The electron at ground state occupy the $1s$ state with orbital angular momentum $L=0$ state. This happens because  $1s$ state is spherically symmetric. However the  intrinsic spin angular momentum of the electron could be quantized in $+z$ or $-z$ direction, this we represents as $| \uparrow \rangle$ and $| \downarrow \rangle$ respectively. When the two Hydrogen atoms are far apart the relative spin states of the two electrons can be shown to be arbitrary. However when the two hydrogen atoms come nearby then the electronic orbital of two electron overlaps which means that electrostatic potential energy increases. To minimise this electrostatic energy and also to maintain the two electronic wave function to be antisymmetric according to Pauli principle, one can show that the electronic spins forms a singlet, which can be shown to be an eigenstate of the operator $\vec{S}_1 \cdot \vec{S}_2$ where
$\vec{S}_{1/2}$ represents the spin angular moments of the electrons. In a nutshell, the origin of magnetic interaction is due to the interplay of minimization of electrostatic energy due to overlapping orbitals in conjunction with the antisymmetrization of wave function according to Pauli principle. Extending this idea to complex condensed matter system composed of varieties of atoms/molecules, different complex spin Hamiltonian can emerge. The most well known of them is so called Heisenberg Hamiltonian $H= \sum_{<ij>}  J_{ij} \vec{S}_i \cdot \vec{S}_j$.  Here $<ij>$ denotes a pair of sites which could be nearest 
neighbour or next nearest neighbour. The sign of $J_{ij}$ could be positive or negative depending on the relative orbital properties of the participating atoms/molecules. With this prelude we describe the Hamiltonian  that is celebrated as Kitaev model. To understand that we first note that for a hexagonal lattice, there are three different kind  of bond that can be grouped according to their alignation as showin in Fig.~\ref{fig1}. There are  vertical bonds and lets call it $z$-bond. There are  bonds with  positive slope and lets call it $x$-bond and the bonds which are having negative slope are termed as $y$-bonds. Now it is clear that a given site is connected to three other sites by these three different type of bonds. The spin-spin interaction between two neighbouring spins depends on the kind of bonds they are connected with. For example, a given spin located at site `$i$' interact with another spin located at  site $i+\delta_x$ with $S^x_i S^x_{i + \delta_x}$, where as the same spin interact with its neighbouring site located at site $i + \delta_y$ and $i + \delta_z$ with $S^y_i S^y_{i + \delta_y}$ and $S^z_i S^z_{i + \delta_z}$ respectively. In the above $S^{\alpha}_i, ~\alpha=x,y,z$ denotes the spin component of the spin angular moment. With this in mind the Hamiltonian for the system can be written as follows, 

\begin{eqnarray}
H= \sum_{<ij>_x} J_x S^x_i S^x_j + \sum_{<ij>_y} J_y S^y_i S^y_j + \sum_{<ij>_z} J_z S^z_i S^z_j
\label{ham0}
\end{eqnarray}

In the above equation the coupling parameter $J_x, J_y, J_z$ are taken positive. However the analysis we will follow and the details of the property that we will discuss
remains unchanged for any combinations of signs of the coupling strength. This is another remarkable characteristics of Kitaev model.

In this pedagogical article we try to give a  few exact results and exciting  physics that this Hamiltonian contains.  To get insight what does this innocently looking Hamiltonian  as given in Eq.~\ref{ham0} contains,  we take simple limits and readers are asked to find the answers  themselves before proceeding further.

\begin{itemize}
\item{\it{Consider the spins as classical spins and find the classical configurations of spins that minimizes the Hamiltonian in the limits a) $J_x=J_y=0, J_z \neq 0$, b) $J_x \neq 0, J_y \neq 0, J_z=0 $ and c) $J_x \neq 0, J_y \neq 0, J_z \neq 0$.}}
\item{\it{For more on classical solution the reader is requested to have a look at the reference by \cite{sen-shankar}}}. 
\end{itemize}

\subsection{Exact Solution of Kitaev model}
\label{section-2-0}

The Hamiltonian is an effective representation of how the constituent particles or spins in question interact and there are energy cost for each specific interactions.  For interacting particles, there are varying degrees of interactions associated with different energy cost. The Hamiltonian is an operator form of energy and its spectrum or possible values of energy that it can have gives a very important physical insight of the system.Thus our primary aim will be to  find the eigenvalues  and eigenfunction of Eq.~\ref{ham0}. Once we have that, we can calculate any physical quantity of our interest in principle.  We are familiar with eigenvalue problem of hydrogen atom, harmonic oscillator, particle in a box etc \cite{house}. Usually this involves solving certain differential equations. For the Hamiltonian in Eq.~\ref{ham0}, individual spins are represented by Pauli matrices which has dimensions two. The individual state of a given spin-1/2 particle can be represented by the two eigenstates of the Pauli matrix.  If there are `$N$' such spin-1/2 objects, the total  dimensions becomes $2^N$ and for this problem it becomes the dimension of Hilbert space which  is $2^N$. The Hamiltonian, in Eq.~\ref{ham0} can be represented  suitably in $2^N \times 2^N$ dimensional hermitian matrix. To obtain the eigenvalue and eigen function one has to diagonalize this matrix. However given the modern day computer which is commonly available, one can solve a system of 30-40 spins. The system with such small number of spins often misses to capture the long range order and the true characteristics of the model developed  only in the thermodynamic limit. For one dimensional system this might be enough. This fundamental limitation is the main challenge to modern condensed matter  researcher.  One commonly used technique to deal with the spin-Hamiltonian is that, we use suitable auxiliary variable to express the spins \cite{tsvelik}. Many times we use fermionc or bosonic field operators to express the spin-1/2 angular momentum algebra. This process is called fermionisation or bosonisation depending on the auxiliary field operator that are used. Below we provide two such very commonly used fermionisation process implemented for one dimensional spin problem and  reader is asked to implement it themselves to understand the usefulness of it.

\begin{center}
  \Large{Exercise-1}
\end{center}
\begin{itemize}
\item{\it{Jordan --Wigner Fermionisation and its application to transverse field Ising Model \cite{jordan,pfeuty}, 1d X-Y model \cite{lieb1} }}
\item{\it{Spin-1/2 one dimensional Kitaev model \cite{1dkm}}}
\item{\it{Apply the Jordan-Wigner Fermionisation for 1d Heisenberg model and explain the differences with the transverse field Ising model, 1d X-Y model or 1d Kitaev model. Compare the results with the one dimensional antiferromagnetic Heisenberg chain \cite{bethe}.}}
\end{itemize}

The Hamiltonian in Eq.~\ref{ham0} is special in the sense that usually the spin-spin interaction between the spins have all the component involved with the expression $S^x_1S^x_2+ S^y_1S^y_2 + S^z_1S^z_2$. However for the Kitaev interaction only one component of spins is involved with a given spin. For this reason Kitaev interaction is also called an anisotropic Heisenberg interaction.  Another key feature of this model is the presence of large number of conserved quantities. For each plaquette we can define an operator  which commutes with the Hamiltonian.  We call this plaquette operator as $B_p$ where the subscript `$p$' stands for the plaquette index. Plaquette operators defined on different plaquettes commute among themselves. With reference to the Fig.~\ref{fig1} $B_p$ is defined as,
\begin{equation}
\label{bpeq}
 B_{p}=\sigma^{y}_{1}\sigma^{z}_{2}\sigma^{x}_{x}\sigma^{y}_{4}\sigma^{z}_{5}\sigma^{x}_{6}. 
\end{equation}
 Thus for each plaquette  `$p$' we can define  a $ B_{p}$. It can be easily checked that,
\begin{equation}
[B_p,H]=0 \, , \,\,\,\, [B_p,B_q]=0, ~p \ne q, 
\label{conbpch1}
\end{equation}
In the above $p,q$ indicate different plaquette indices. This implies  that $B_p$'s are  conserved quantities for this model.  It is easy to verify that $B^2_p=1$  which implies that eigenvalues of $B_{p}$ are $\pm 1$.  We will see later that this conserved quantity plays a significant role in the dynamics of Kitaev model. We now present the formal solution of this spin model as obtained by  Kitaev himself \cite{kitaev-2006}. He showed that this  spin model can be solved exactly using a fermionisation procedure which expresses the spin 1/2 operators in terms of Majorana fermion operators. In the next  we elaborate on this.

\subsection{\label{fso2.1}Fermionisation of spin 1/2 operators}

Apart from the Fermionisation procedure mentioned in the exercises in Sec.~\ref{fso2.1}, here we outline another fermionisation procedure which was used by Kitaev  himself. We again ask the reader to do the following exercize before proceeding further.

\begin{center}
  \Large{Exercise-2}
\end{center}

\begin{itemize}
\item{\it{Define the operator, $c=(c_{1}+c^{\dagger}_{1}), c^{x}=\frac{1}{i}(c_{1}-c^{\dagger}_{1}),~ c^{y}=(c_{2}+c^{\dagger}_{2}),~c^{z}=\frac{1}{i}(c_{2}-c^{\dagger}_{2})$. Verify that they satisfy usual anticommutation relation, they are self conjugate and when multiplied with itself gives unity.}}
\item{\it{Define $\sigma^{x}=ic^{x}c,~\sigma^{y}=ic^{y}c,~\sigma^{z}=ic^{z}c$. Verify that Pauli matrices satisfy usual commutation relation.}}
\item{ \it{Check whether the above definition reproduces $ \sigma^x \sigma^y \sigma^z=  \pm i$ for the odd particle states and even particle states respectively.}}
\item{\it{Find the physical Hilbert space corresponding to the definition. Remember that a spin-1/2 particle has a matrix representation by Pauli matrices and has  a Hilbert space dimension two. However in the auxiliary variable when we express the spin variable by fermionic operators, we used two fermions. Now each fermions has a Hilbert space dimension two, thus two fermions together has Hilbert space dimension of four. Thus the initial physical Hilbert space of a given spin of dimension  two has been mapped to a Hilbert space dimension of four. This causes enlargement of Hilbert space dimensions and accounts for unphysical states. The previous exercise is aimed at having an idea about sub-classification of this extended Hilbert space where $ \sigma_x \sigma_y \sigma_z=  \pm i$ needs to be worked out.}}
\end{itemize}

\subsection{\label{qdham1}Quadratic Hamiltonian}
To proceed we  write the Hamiltonian in terms of fermionic operators as discussed above. After inserting the relations given  in Exercise-2 in  Eq.~\ref{ham0}(i.e $\sigma^{\alpha}_i= i c^{\alpha}_i c_i,~~\alpha=x,y,z$), it reduces to

\begin{eqnarray}
\label{2}
 H &=&\sum_{x-{\rm link}}J_{x}(ic^{x}_{i,a}c^{x}_{j,b})ic_{i,a}c_{j,b}+
\sum_{y-{\rm link}}J_{y}(ic^{y}_{i,a}c^{y}_{j,b})ic_{i,a}c_{j,b} \nonumber \\
&+& \sum_{z-{\rm link}}J_{z}(ic^{z}_{i,a}c^{z}_{j,b})ic_{i,a}c_{j,b}
\end{eqnarray}
In the above equation `$a$' and `$b$' denotes the sublattice index.
We observe that each term in the above Hamiltonian is quatric in  Majorana fermion operators. Generally such quatric Hamiltonian is quite difficult to solve. However it can be easily noted that  operators in the parenthesis of each term of the above Hamiltonian  commute with the Hamiltonian and commute among themselves.  Which means they follow the following commutation relation which the interested readers are requested to check before proceeding further.
\begin{eqnarray}
[ic^{\alpha}_{i,a}c^{\alpha}_{j,b}, ic^{\beta}_{i,a}c^{\beta}_{j,b} ]=0, [ic^{\alpha}_{i,a}c^{\alpha}_{j,b}, ic_{i,a}c_{j,b} ]=0
\end{eqnarray}

In the above equation $\alpha, \beta=x,y,z$ and $(i, j)$ denotes a pair of nearest neighbour sites joined by a $\alpha$-type bond. It means that they are conserved quantities as far as this fermionised Hamiltonian is concerned. This fact makes Eq.~\ref{2} to be effectively quadratic in Majorana fermions. Let us call, $ic^{x}_{i,a}c^{x}_{j,b}=u^{x}_{i,j}$ for the $x$-link. Similarly we define $ u^{y}_{i,j}$ and $ u^{z}_{i,j}$ on $y$ and $z$ links respectively. It is obvious that, $u^{\alpha}_{i,j}=-u^{\alpha}_{j,i}$ and its eigenvalues can take value $\pm 1$ given the fact $(u^{\alpha}_{i,j})^2=1$.  Here we follow the convention of keeping the indices of the site belonging to the `$a$' sub lattice first and then for `$b$' sub lattice in the expression of $u_{i,j}$(fro brevity many times explicit mention of $\alpha$ in the expression of $u^{\alpha}_{ij}$ will be avoided.). 

\begin{center}
  \Large{Exercise-3}
\end{center}
\begin{itemize}
\item \it{ Using the definition of fermionization in Exercise-2 and definition of $B_p$ as in Eq.~\ref{bpeq} show that }\\
\begin{equation}
\label{bpmajo}
 B_{p}=\prod_{(j,k)\epsilon {\rm boundary(p)} } u^{\alpha}_{j,k}.
\end{equation}
\end{itemize}

Using the above replacements for the conserved quantities on each bond, the Hamiltonian takes the following form,
\begin{eqnarray}
H &=&\sum_{<ij>_x} J_{x}u^{x}_{i,j}ic_{i,a}c_{j,b}+ \sum_{<ij>_y}
J_{y}u^{y}_{i,j} ic_{i,a}c_{j,b}\nonumber \\
&&  +\sum_{<ij>_z}J_{z} u^{z}_{i,j}ic_{i,a}c_{j,b} \\
&& = \Psi^{\dagger}  \mathcal{H}([u]) \Psi
\label{3}
\end{eqnarray}
where $ \Psi^{\dagger}=(c_1,c_2,...,c_N)$ is an one dimensional row matrix of length `$N$' and $\mathcal{H}[u]$ is a $N \times N$ matrix which depends on the values of $u_{ij}$ on each link. Now we see that above Hamiltonian describes a tight binding Majorana fermion hopping interactions but the hopping matrix elements are  coupled with  conserved operator or fields $ u^{\alpha}_{i,j}$ on each bonds. As we mentioned that this operators have eigenvalues $\pm 1$. These $u^{\alpha}_{i,j}$ are called $Z_2$ gauge fields because of the eigenvalue to be $\pm$. Depending upon the values of these $Z_2$  gauge fields the eigenvalues of the system will change. Physically one way to visualise is that initially we had quartic Majorana fermion interaction of the form $ic^{\alpha}_{i,a} c^{\alpha}_{j,b} i c_{i,a} c_{j,b} $. This process can be visualise in many ways. For example this can be thought   as two Majorana fermion hopping process happening simultaneously such as $c^{\alpha}_{i,a} \rightarrow c^{\alpha}_{j,b},~c_{i,a} \rightarrow c_{j,b}$ or $c^{\alpha}_{\alpha,i,a} \rightarrow c_{j,b} ,~c_{i,a} \rightarrow c^{\alpha}_{j,b}$ or similar processes. Alternatively it can be thought of  as two on site interactions happening simultaneously  yielding an energy cost. This on sight interaction involves the Majorana fermions at a given site only. All possible equivalent explanation exists for such four body interactions and they all are true.  However the fact that the combination $ic^{\alpha}_{i,a} \rightarrow c^{\alpha}_{j,b} $ is a conserve quantity signify that once  eigenvalue of $ic^{\alpha}_{i,a} \rightarrow c^{\alpha}_{j,b}$ is fixed to either +1 or -1 (like an initial condition) it remains same with time. Lets assume that the eigenvalue is fixed at +1(or -1) then the hopping process $ c_{i,a} \rightarrow c_{j,b}$ with an phase +1( or -1). This also signify that the process $c^{\alpha}_{i,a} \rightarrow c_{j,b} ,~c_{i,a} \rightarrow c^{\alpha}_{j,b}$ or other possible processes are not allowed by the system. Further in Eq.~\ref{3}, if we transform $c_{i,a} \rightarrow \lambda_i c_{i,a} $, we observe that to keep the Hamiltonian invariant under such transformation $u^{\alpha}_{i,j}$ must change according to $u^{\alpha}_{i,j} \rightarrow \lambda_i u^{\alpha}_{i,j} \lambda_i$, where the allowed value of $\lambda_i$ is again $\pm 1$. THus the Hamiltonian given in Eq.~\ref{3} has a underlying $Z_2$ symmetry \cite{wen}.
\begin{center}
\Large{Exercise-4}
\end{center}
\begin{itemize}
\item{\it {Derive the Hamiltonian as in Eq.~\ref{3}}}
\item{\it{Prove that $[H, u^{\alpha}_{ij}]=0,~~~[u^{\alpha}_{ij},u^{\beta}_{kl}],~~\alpha,\beta=x,y,z$}}
\item{\it{Derive the Heisenberg equation of motion for the operator $i c_{i,a} c^{\alpha}_{j,b} $ and $i c^{\alpha}_{i,a}c^a_i $ and try to explain it with your own understanding.}}
\end{itemize}

   It is to be noted that the new conserved quantities, $u_{i,j}$, were absent in the original spin Hamiltonian. In the original spin Hamiltonian as given in Eq.~\ref{ham0}, there is no conserve quantity associated with a bonds. Earlier we found that there was only one type of conserve quantity associated with each hexagonal plaquette only and it is given in Eq.~\ref{bpeq}. Also from exercise 3 (Eq.~\ref{bpmajo}), we found that the $B_p$ is a product of six  conserved quantities ($u_{i,j}$'s) defined on each bond. Initially our physical degrees were the spins represented by a product of two Majorana fermions. The bond conserved quantity $u_{i,j}$ is also product of two Majorana fermions but each Majorana fermion is taken from the two spins attached at the end of a particular bond. And one can verify that a $u_{ij}$ can not be expressed in terms of spins. It is the product of $u_{ij}$ over the links of a hexagonal plaquette which can be expressed as a product of six spin operators and called $B_p$. One can easily verify that the square of $B_p$ is one indicating that the eigenvalue of the operator $B_p$ is $\pm 1$. We have also seen that eigenvalue of the bond-conserve quantity $u_{ij}$ is also $\pm 1$. As $B_p$ is expressed as a product of six $u_{ij}$ we can understand that for a given eigenvalue of $B_p$ there are many choices for the eigenvalues of the participating $u_{ij}$'s. Among the total $2^{6}$ configurations that six $u_{ij}$ provides, half of them  yields $B_p=1$ and other half
yields $-1$. For a given  eigenvalue of $B_p$ say +1, among the various combinations, we will observe that a given $u_{ij}$ changes its eigenvalue from +1 to -1 though value of $B_p$ is fixed to 1. For this reason $u_{ij}$s are called gauge fields analogous to magnetic vector potential and $B_p$'s are physical observable analogous to Magnetic field. Physically $u_{ij}$ can not be measured and its  expectation value or outcome in experiment will be zero because of gauge averaging over many combinations that one $u_{ij}$ takes. However eigenvalue of $B_p$ is a physical observable. Thus we will use the phrase that $B_p$ is gauge invariant and $u_{ij}$ are not gauge invariant. \\
\noindent

\begin{center}
\bf{Technicality}
\end{center}

\begin{center}
\begin{figure}
\psfrag{a}{$2^{N}$}
\includegraphics[width=0.6\linewidth]{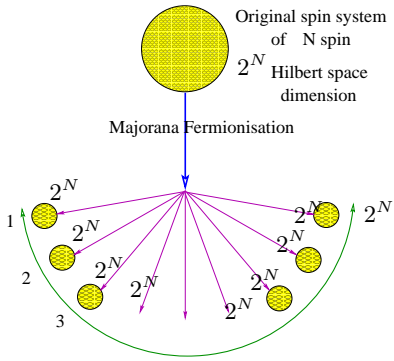}
\caption{\label{hilbert}The big circle at the top denotes the original Hilbert space dimension of $2^{N}$. At the lower the $2^{N}$ copies of the original Hilbert space is shown. This happens due to Majorana fermionisation using four Majorana fermions at a given site. The description of Kitaev model in each of these $2^{N}$ copies are identical up to an gauge transformations which would connect one gauge copy to another.}
\end{figure}
\end{center}

 Before proceeding further we elaborate on the Hamiltonian represented in  Eq.~\ref{3}. The motivation is that physics is not always a story telling. Many truth of a given physical system can be understood by an intelligent mind without going into the details of mathematics. But very often, the situations become so complex for a many body system that one needs to rely on mathematics. In Kitaev model such mathematics become exact beautifully and this motivates us to explain in detail the consequences of mapping of original spin problem into a Majorana fermion hopping problem coupled with $Z_2$ conserved gauge field. Such an exact description may  happen for other system approximately and we believe that a thorough understanding of it in the context of Kitaev model will help the reader to extend their imagination easily to explore the intricacies of other complex system. Lets consider a  finite Honeycomb lattice such that it has $N_1$ dimer or $z$-bond in the $\vec{a}_1$ direction and $N_2$ dimer or $z$-bonds in $\vec{a}_2$ direction. Thus we have a system of $N$ spins with $N=2 N_1 N_2$. Because there are two spins in a given $z$-bonds. Total number of  bonds of a particular type say $x,~y,~z$ are equal and they are $N_1 N_2$. Thus we have a total number of bonds $N_b=3 N_1 N_2=3N/2$. The original spin Hamiltonian of $\mathcal{N}$ spins is defined in $2^{\mathcal{N}}=M$ dimensional Hilbert space. This means that we are to solve a $M \times M$ matrix with $M$ real eigenvalues and eigenstates. Now while we had employed a fermionization procedure where at each site there are two fermions implying that  each site is associated with a Hilbert space dimension of four yielding total Hilbert space of $4^{\mathcal{N}}= (2^2)^N=2^N 2^N=M_1=2^M$ which means we have now enlarged Hilbert space yielding $M_1$ number of eigenstates/eigenvalue which is much more than the original Hilbert space. Thus though the original problem in spin-space was dimension $2^N$, after fermionisation now we have a problem of dimension $2^N \times 2^N$. This situation is depicted in Fig.~\ref{hilbert}. 

\begin{figure}
\psfrag{a}{$N \times N$}
\includegraphics[width=0.5\linewidth]{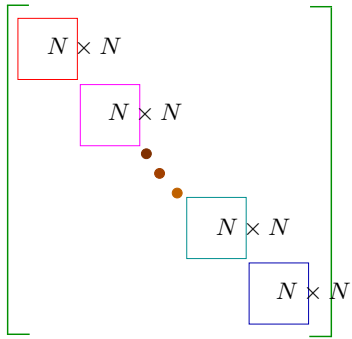}
\caption{\label{matrix} For a system of `$N$' sites we have a matrix of dimension $N \times N$ for a given realisation of $u_{ij}$ on each link. Such matrix is shown by the small coloured matrix. For a given $N$ sites,
there are $3N/2$ bonds yielding a total possibility of $2^{3N/2}$ such configuration. Thus the dimension of the outer big matrix is $2^{3N/2} \times 2^{3N/2}$. For more detail on the counting on the Majorana fermion and its relation to complex fermions see text below. }
\end{figure}

\noindent
    Let us try to find another connection which will explain the relations between the eigenstates of the original spin problem given by Eq.~\ref{ham0} or Eq.~\ref{3} and Majorana fermion hopping problem given by Eq.~\ref{2}. In the spin space we must have $M=2^N$ number of eigenstates and eigenvalues. How do we obtain that from Eq.~\ref{3} and what is the consequences of enlargement of Hilbert space dimension. This will be explained here.  For a given distribution of gauge fields on the bonds of honeycomb lattice, we have a matrix  of dimension $N$ which has $N$ eigenvalues and $N$ single particle Majorana fermion eigenstates.  If one diagonalizes $\mathcal{H}[u]$ in Eq.~\ref{3}, we obtain $\lambda[u]_i, i=1,N$ with eigenvectors $|\lambda[u] \rangle_i= \sum_i \gamma[u]_i c_i$. However these Majorana fermions are to be regrouped to yield $N/2$ complex fermions which have a definite occupation number representation~\cite{bruus}.  From this $N/2$ single particle eigenstates one can obtain $2^{N/2}$ many-body states by constructing eigenstates having arbitrary particle number. The energy eigenvalues for such many body states can be written as $\sum_i \lambda[u]_i$. One can easily note that there are $2^{N/2}$ such many body eigenstate which can be obtained by $\sum_m { \frac{N}{2} \choose m}$. However we remember that such description is true for a given distribution of conserved quantities on every bonds. This situation is depicted in Fig.~\ref{matrix} where the dimension of the outer big matrix is $\frac{N}{2} \times \frac{3N}{2}= \frac{3N^2}{4}$. Inside this big matrix the smaller diagonal matrix represent a certain distribution of  conserved gauge field and the dimension of this smaller matrix is $N$ yielding $N/2$ complex single particle complex fermionic eigenstate which in turn yield $2^{N/2}$ many body state.  Now we already mentioned there are in total $N_b=3N/2$ number of bonds yielding in total $2^{3N/2}$ such combinations. This means, in reference to Fig.~\ref{matrix}, we have  $2^{3N/2}$ such small diagonal matrix. Each of this smaller matrix yields an eigenstates of $2^{N/2}$ number. Thus the total number of eigenstates is $2^{N/2}  2^{3N/2}= 2^N 2^N $ which matches with our earlier counting. \\
\noindent

Earlier we mentioned that $u_{ij}$'s are not gauge invariant quantities and they are not physical observable. The gauge invariant physical observables are the plaquette conserve quantity $B_p$. There are many combinations of $u_{ij}$ which yields a configurations of $B_p$ for each plaquette. And it is important to note that the energy eigenvalue only depends on the distribution of $B_p$ not $u_{ij}$. If two configurations of $u_{ij}$ yields same distributions of $B_p$, they should have identical eigenvalues. In reference to Fig.~\ref{matrix}, this means that there are many gauge equivalent smaller matrix which gives identical distributions of $B_p$ and have identical energy eigenvalues. We leave it to the interested reader to check themselves with the following exercise.

\begin{figure}
\psfrag{A}{A}
\psfrag{B}{B}
\psfrag{C}{C}
\psfrag{D}{D}
\psfrag{+}{+}
\psfrag{-}{$-$}
\psfrag{1}{1}
\psfrag{2}{2}
\psfrag{3}{3}
\psfrag{4}{4}
\psfrag{5}{5}
\psfrag{6}{6}
\psfrag{7}{7}
\psfrag{8}{8}
\includegraphics[width=0.8\linewidth]{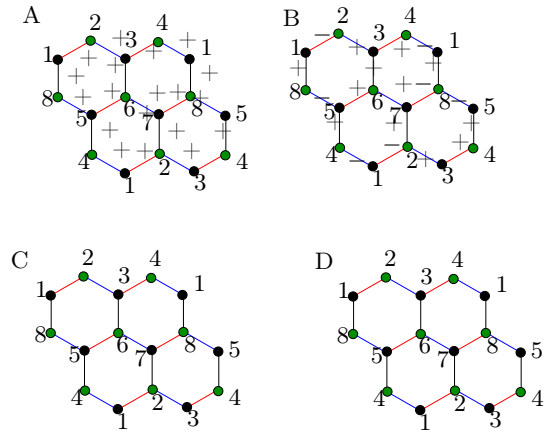}
\caption{\label{fourchoice}In the panel A and B, two choices of conserved $Z_2$ gauge fields are shown to yield identical distribution of $B_p$. For C and D two equivalent $Z_2$ gauge field choices are to be found to yield $B_p=-1$ for all the plaquette.}
\end{figure}

\begin{center}
\Large{Exercise-5}
\end{center}
\begin{itemize}
\item{{ \it Calculate the number of ways in which one  can have $B_p=1$ for all the plaquette. Is this number same for having a random configuration of $B_p$ for all the plaquette.}}
\item{{\it In Fig.~\ref{fourchoice}, A and B we have given two configurations for $B_p$=1 for all the plaquette. Following the procedures of Ref.~\cite{kitaev-2006}, Sec-5, construct the eigenvalues and check that they yields identical spectrum. Construct similar distributions of gauge field such that it gives $B_p=-1$ for all the plaquette. Calculate the spectrum as done for $B_P=1$ and find which configuration has minimum ground state energy. Explain your finding. }}
\end{itemize}
\vspace{0.5cm}
\begin{center}
\Large{Lieb Theorem }
\end{center}

We observe from Eq.~\ref{3} that Hamiltonian is functional of configurations of conserve quantity $u_{ij}$ defined on each bonds. Such Hamiltonian can be represented by block diagonal form in the eigenstate of $u_{ij}$ as represented by the small square matrix in Fig.~\ref{matrix}. Each sub-block refer to a certain distribution of $u_{ij}$ and we also explained before that  each block corresponds to a certain distribution of gauge invariant conserve quantity $B_p$ for each plaquette. There are many distinct configurations of $u_{ij}$ which yields a unique configurations of $B_p$. We also know that  eigenvalue of $B_p$ is $\pm 1$. Now the important question is the following. Does the  ground state obtained from each sub-square block in Fig.~\ref{matrix} yields same energy or does it depends on the distribution of $u_{ij}$ or does it depends on the distribution of $B_p$. As the $u_{ij}$ is not a gauge invariant object and physically not observable, the energy is not directly dependent on the distribution of $u_{ij}$ rather it depends on the distribution of $B_p$. The next question is which distribution of $B_p$ yields the absolute minima. From a very remarkable theorem \cite{lieb} by E. Lieb. we know that uniform configuration of $B_p=1$ for each hexagonal plaquette yields the global minima. This is obtained by fixing $u_{i,j}=1$ for every link (there are many other configurations which yields $B_p=1$, however each of them would yield identical ground state energy). This has also been confirmed by Kitaev numerically \cite{kitaev-2006}. For the uniform choices of $u_{i,j}$(which corresponds to global minima ) we can easily diagonalise the Hamiltonian and get the ground state wave function.

\begin{center}
\Large{Physical wave function and projection operator}
\end{center}

With the discussion of the foregoing paragraph, let us call the wave function obtained by diagonalizing  Eq.~\ref{3} with uniform configuration of $u_{ij}=1$(yielding $B_p=1$ for each hexagonal plaquette) as $|\psi\rangle_{ext}$. We have deliberately added the subscript `$ext$'  to remind the fact that the above wave function is obtained  in the extended Hilbert space. One may question whether the  ground state energy obtained in such a way is the true ground state energy which should have been obtained in the physical Hilbert space. Moreover what about the wave function it self?  Because to calculate other physical observable we need the true ground state belonging to the physical Hilbert space. Otherwise $|\psi\rangle_{ext}$ is not of much useful. Now it is time to discuss this issue.  Whenever we make an operation which takes us from an Hilbert space $H^{>}$ with more number of states to another Hilbert space $H^{<}$ with less number of states such that some states are excluded in $H^{<}$, we need an projection operator. If any states $\Psi^{>}$ belongs to $H^{>}$, the corresponding states in $H^{<}$ after projection is obtained as
$\Psi^{<}= P \Psi^{>}$ where $P$ is the projection operator. The job of the projection operator is to remove the unphysical states and keep only the physical state in the expansion of $\Psi^{>}= \gamma_i  |i \rangle $ where $\gamma_{i}$ is complex coefficient and $|i \rangle$ is the normalized basis vector belonging to $H^{>}$. Actually $| i \rangle$ can be decomposed into two groups $| i_< \rangle$ and $| i_> \rangle$ such that $ | i_< \rangle$  constitute the normalized  basis  vector of $H^{<}$. The job of $P$ is to remove or annihilate the states $| i_> \rangle$ such that $P \Psi^{>}$ involves only the states belonging to $H^{<}$. 
\begin{equation} 
\label{4}
|\psi\rangle_{\rm phy}=\hat{P}|\psi\rangle_{ {\rm ext}}
\end{equation}
\indent 

Now we  try to understand the projection operator and its expression in the context of Majorana fermionisation in a little depth and we  find a useful connection among the various copies of the Hilbert space that we mentioned earlier. Lets $| \uparrow \rangle$ and $ | \downarrow \rangle$ be the eigenstates of $\sigma_z$ with eigenvalue $\pm 1$. The action of $\sigma_x$ and $\sigma_y$ on these
two states are defined as $\sigma_x |\uparrow (\downarrow) \rangle = | \downarrow (\uparrow) \rangle$ and $\sigma_y |\uparrow (\downarrow) \rangle = i(-i)| \downarrow (\uparrow) \rangle$. Now the complex fermion $c_i, c^{x}_{i,a},c^{y}_{i,a},c^{z}_{i,a}$ that has been used
 to define the Majorana fermionisation of spin operators has the states $|00 \rangle, ~|10 \rangle= c_1^{\dagger}|00 \rangle , ~|01 \rangle= c_2^{\dagger}|00 \rangle,~ |11 \rangle =c^{\dagger}_1c_2^{\dagger}|00 \rangle$. It is clear that the original spin  states needs to be mapped to these four states and there is an enlargement of states. To understand the mapping let us recall that $D=\sigma_x \sigma_y \sigma_z=i$ is an identity which must be hold for any states. If we calculate $D$ according to the definition  given in Exercise-2 one finds $D=i (1-2 c^{\dagger}_1 c_1) (1-2 c^{\dagger}_2 c_2)$. Now we see that $D=i$ holds true only for the states $|00\rangle$ and $|11\rangle$. Thus any physical states should  have the general representation $\Psi_{ph}= a_{00}|00\rangle  + a_{11} |11\rangle$. But while working on the extended Hilbert space we  encounter states $\Psi_{ext}= a_{00}|00\rangle  + a_{11} |11\rangle +a_{01}|01\rangle  + a_{10} |10\rangle$. How do we get rid of the unphysical states $|01 \rangle$ and $|10 \rangle$. Note that the operator $P=(1+D)/2$ acting on this unphysical states yields zero and keeps the physical states as it is. Thus it is straightforward to check that $P \Psi_{ext}= \Psi_{ph}$. Now this is about  a given site and we call this projection
operator $P_i= (1+ D_i)/2$. The above explanation can be extended to all the sites of the  entire system and the total projection operator is defined as,  
\begin{equation}
\label{5}
\hat{P}=\prod_{i \epsilon {\rm all~~sites}}\frac{(1+D_i)}{2} 
\end{equation}

Before, we go for a formal solution we advise the reader to go through the following exercise which will strengthen
their familiarity with the concept of Majorana fermionisation.

\begin{center}
\Large{Exercise-6}
\end{center}

\begin{itemize}
\item{\it{Find the eigenstates of the operator $c, c^x, c^y, c^z$ defined in Exercise 2.}}
\item{\it{Find the eigenstates of the operator $\sigma^x, \sigma^y, \sigma^z$ in the Fock
space of four states described above and see the consequences of projection operators on
them.}}
\end{itemize}

\subsection{The ground state}
\label{grst}
We have already argued that it is the uniform configuration of $B_p=1$ which contains the global
minima of the spectrum. Here we consider the choice $u_{ij}$ =1 for each link which is one of
the realizations of $B_p=1$ for each plaquette. After doing that the  Majorana fermion
hopping Hamiltonian given in Eq.~\ref{ham0} reduces to a translational invariant Hamiltonian facilitating easy solution
using Fourier transformations. The translational invariant Hamiltonian  is given by,

\begin{eqnarray}
\label{ham}
 H &&=\sum_{x-{\rm link}}J_{x}ic_{i,a}c_{j,b}+
\sum_{y-{\rm link}}J_{y}ic_{i,a}c_{j,b} + \nonumber \\
 &&~\sum_{z-{\rm link}}J_{z}ic_{i,a}c_{j,b}
\end{eqnarray}
 To solve the above Hamiltonian we define the following Fourier transformations for the Majorana fermions, 
 \begin{equation}
 c_{i,a(b)}= \sum_{k}\frac{1}{\sqrt{MN}} e^{i\vec{k}.\vec{r}} c_{k,a(b)}.
 \end{equation} 

Here we have taken a lattice with $M$ and $N$ unit cells in the directions of $\vec{a}_1$ and $\vec{a}_2$ respectively as shown in Fig.~\ref{fig1}. Here  $\vec{r}= m \vec{a}_1 +n \vec{a}_2$ and ${\vec{k}}=\frac{p}{M}{\vec{b}_1}+\frac{q}{N}{\vec {b}_2}$, where ${\vec b_{1,2}} $ are the reciprocal lattice vectors are given by,
\begin{equation}
\label{revec}
{\vec{b_{1}}}= \frac{4\pi}{\sqrt{3}}(\frac{\sqrt{3}}{2} \textbf{e}_{x} +\frac{1}{2} \textbf{e}_{y}) \,\,;\,\, {\vec{b_{2}}} = \frac{4\pi}{\sqrt{3}}  \textbf{e}_{y} 
\end{equation}
 Here `$p$' and `$q$' varies from $-M/2$ to $M/2$ and $-N/2$ to $N/2$ respectively. The above discussion defines the Brillouin zone. We  notice  that the property $c^{\dagger}_i=c_i$  implies $c_{k}=c^{\dagger}_{-k}$. After performing the Fourier transformation, we get the Hamiltonian in k-space as follows,
\begin{eqnarray}
\label{k-hamch2}
H&=& \sum_{k \epsilon {\rm HBZ}} {(c^{\dagger}_{k,a}   c^{\dagger}_{k,b})} \left( \begin{array}{rr} 
0& if^{*}_{k}\\
-if_{k}&0 \end{array} \right) \left( \begin{array}{rr} 
c_{k,a}\\
c_{k,b} \end{array} \right)
\end{eqnarray}
In the above equation `HBZ' stands for half Brillouin zone. Note that for the condition $c_k=c^{\dagger}_{-k}$, all the Majorana modes are not independent. Their is a specific way a Majorana fermion  with positive momentum is related with a Majorana fermion with a negative momentum with the same magnitude. Equivalently  the $c_k=c^{\dagger}_{-k}$  implies  that annihilating a Majorana fermion with momentum $k$ is same as creating a Majorana fermion with momentum $-k$. The spectral function $f_{k}$ is given by,
\begin{equation}
\label{spectral}
f_{k}= J_{z} + J_{x} e^{-ik_{1}} + J_{y} e^{-ik_2}
\end{equation}
In above expression `$k_{1}$' and `$k_2$' are the components of $\vec{k}$ along `$x$' bond and `$y$' bond respectively. They are given by,
\begin{center}
\begin{figure}
\includegraphics[width=0.6\linewidth]{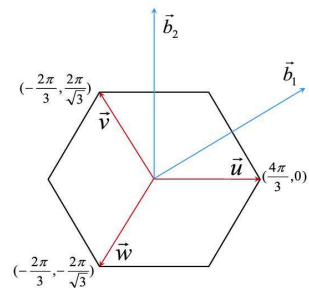}
\caption{\label{bzone}In the above we have plotted the first Brillouin zone for the honeycomb lattice $\vec{b}_1$ and $\vec{b}_2$ denotes the reciprocal lattice vector as defined in Eq.~\ref{revec}.}
\end{figure}
\end{center}

\begin{eqnarray}
k_1&=&\vec{k}.\hat{n}_1 \,\,;\,\, k_2=\vec{k}.\hat{n}_2 \\
\hat{n}_1&=& \frac{1}{2}\hat{e}_x +\frac{\sqrt{3}}{2} \hat{e}_y \,\,;\,\,\hat{n}_2= \frac{-1}{2}\hat{e}_x +\frac{\sqrt{3}}{2} \hat{e}_y
\end{eqnarray}
Here $\hat{n}_1$ and $\hat{n}_2$ are the unit vector along the `$x$' and `$y$' bond respectively. The Hamiltonian given in Eq.~\ref{k-hamch2}  can be diagonalised easily with the following unitary  transformation given below,
\begin{eqnarray}
\label{kham1}
\left( \begin{array}{r} c_{k,a}\\c_{k,b} \end{array} \right) = \frac{1}{\sqrt{2}} \left( \begin{array}{rr} v_{k} &-v_{k} \\
1&1\end{array} \right)  \left( \begin{array}{r} \eta_{k} \\ \xi_{k} \end{array} \right),
\end{eqnarray}

with $v_{k}=i f^{*}_{k}/ |f_{k}|$. The diagonalised Hamiltonian is given by,
\begin{eqnarray}
\label{h-diag}
H= \sum_{k} E_{k} (\eta^{\dagger}_{k} \eta_{k} - \xi^{\dagger}_{k} \xi_{k}),
\end{eqnarray}
where $E_{k}=|f_{k}|$ is the quasi particle energy associated with new field operators $\alpha_k$ and $\beta_k$. The ground  state is obtained by filling up all the negative energy states of  $\beta_k$ quasi particles and can be written as,

\begin{equation}
\label{gr-st}
|G \rangle = \Pi_{k, {\rm HBZ}} \beta^{\dagger}_{k} |0 \rangle,
\end{equation}

where $ |0 \rangle$ represents the  vacuum state such that $\alpha_{k} |0 \rangle = \beta_{k} |0 \rangle =0$. Here the summation is over the half Brillouin zone as explained before. At this point it is important whether the spectrum is gapped or not which implies that if we want to create an excitation over the ground state do we require   finite energy or not. If $E_k=0$ for some values of `$k$' we need no finite energy to create an excitation over the ground state and the system is called  gapless system. Whether a system is gapless or not has much bearing to the thermodynamic quantities of the system as it determines how the one part of the system responses due to disturbances at some other part. To find whether the spectrum is gapless or not we solve for  $E_k=0$ which implies $f_k=0$. It turns out that the, $f_k=0$ has solutions if and only if $|J_x| ,|J_y|, |J_z|$ satisfy the following triangle inequalities:

\begin{equation}
\label{phasecon}
|J_x| \le |J_y| +|J_z| , \,\,\, |J_y| \le |J_x| +|J_z| , \,\,\, |J_z| \le |J_x| +|J_y|
\end{equation}
\begin{center}
\begin{figure}[h!]
\psfrag{x}[lb][lb][1]{$J_x=0$}
\psfrag{y}[lb][lb][1]{$J_y=0$}
\psfrag{z}[lb][lb][1]{$J_z=0$}
\includegraphics[width=0.5\linewidth]{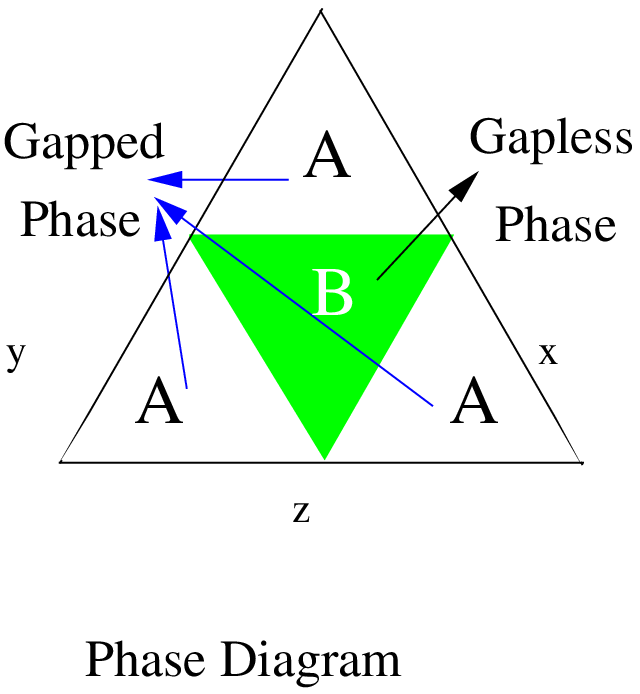}
\caption{\label{phased}Phase diagram for Kitaev model in the parameter space. A point in the above triangle describes relative magnitudes of $J_x, J_y, J_z$. Three sides of the triangle describe $J_x=0, J_y=0$ and $J_z=0$ as given in the figure. The region `A' is gapped and the region `B' is gapless. The gapless region acquires a gap in the presence of Magnetic field.}
\end{figure}
\end{center}
The above inequalities as given in Eq.~\ref{phasecon} can be represented as a point inside a equilateral triangle which has been shown in Fig.~\ref{phased}.

In Fig.~\ref{phasegapless} and Fig.~\ref{phasegapped}, we plotted how the spectrum looks like for gapless and gapped phase respectively.
\begin{center}
\begin{figure}
\psfrag{d}{-2}
\psfrag{z}{$E_k$}
\psfrag{b}{2}
\psfrag{x}{$k_x$}
\psfrag{y}{$k_y$}
\psfrag{c}{-5}
\psfrag{a}{0}
\psfrag{e}{5}
\includegraphics[width=0.8\linewidth]{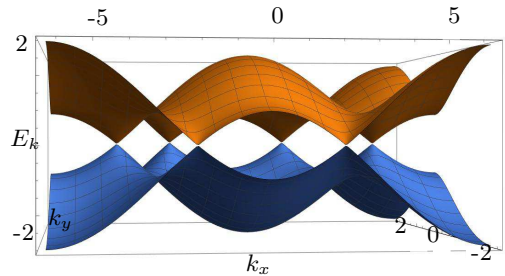}
\caption{\label{phasegapless} Spectrum $E_{k}$ as defined in Eq.~\ref{h-diag} has been plotted above for $J_x=J_y=J_z=1$. We observe that there are six points in the Brillouine zone where $E_k$ vanishes resulting a gapless spectrum. The dispersion near this gapless points are also linear.}
\end{figure}

\begin{figure}
\psfrag{d}{-4}
\psfrag{z}{$E_k$}
\psfrag{b}{2}
\psfrag{x}{$k_x$}
\psfrag{y}{$k_y$}
\psfrag{c}{-2}
\psfrag{a}{0}
\psfrag{e}{4}
\psfrag{f}{-5}
\psfrag{g}{5}
\includegraphics[width=0.75\linewidth]{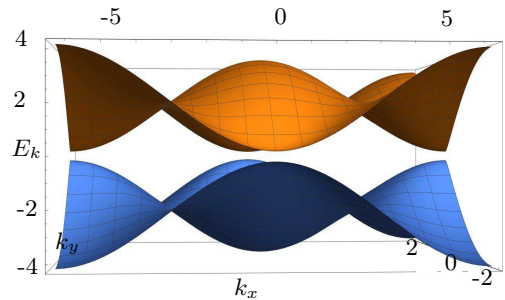}
\caption{\label{phasegapped}In the above $E_{k}$ has been plotted for $J_x=J_y=1,~ J_z=2.2$. We note that there is a gap between  valence band and conduction band. The spectrum near the minimum for conduction band is quadratic.}
\end{figure}
\end{center}

If the inequalities are strict, there are exactly two solutions: ${\bf k}=\pm {\bf q_*}$, one in each HBZ . The region defined by inequalities in Eq.~\ref{phasecon} is the shaded region  B in Fig.~\ref{phased}; this phase is gapless. The region marked by A is gapped. The low energy excitations are different in these two phases. In the gapless phase the low energy excitations are the Majorana fermions but in the gapped phases the low energy excitations corresponds to the vertex excitations which corresponds to the excitations of $B_p$, the conserved quantities. In the presence of magnetic field the phase B acquires a gap. These two regions are topologically distinct as indicated by spectral Chern number which is zero for phase A and one for the phase B  \cite{kitaev-2006}. We have argued that as the projection operator $\hat{P}$ commutes with the Hamiltonian, the solution obtained in the extended Hilbert space is exact. One can indeed show  that there  exist a non zero projections. But this method of solving gives eigenstates of the Hamiltonian as well as the eigenstates  of the conserved quantities in terms of Majorana fermions whose occupation number is not well defined. In the next section, we would extend the Majorana fermionisation in  an useful way such that  eigenfunctions of the Hamiltonian as well as $u^{\alpha}_{ij}$  is represented by the usual  occupation number representation of complex fermions \cite{bruus}.


\section{Order parameter}
\label{sec-order}

We have seen that the initial spin-Hamiltonian  given in Eq.~\ref{ham0} is reduced to an effective  quadratic fermionic Hamiltonian as given in Eq.~\ref{2}. The initial physical object was spin 1/2  magnetic moment. For a two spins belonging to two different sites, spin-angular momentum  commute with each other if we express them by suitable representation. For spin-1/2 particle, the Pauli matrices are a faithful representations. However, the effective Fermionic Hamiltonian consists of Majorana fermions which aniticommutes. This conversion of effective degrees of freedom is  known as emergent degrees of freedom due to interactions. However there is a definite connections between them.  The  eigenstates either obtained by diagonalizing the  original spin-Hamiltonian or the effective Majorana Hamiltonian has one to one correspondence and they have identical eigenvalue spectrum. In  dealing with physical systems, generally one is interested with the ground state wave functions at low temperature.  Now we must ask  what characterizes the ground states, what is the properties of the ground states of different phases, that will differentiate one phases from another. various order parameters are used to determine a certain phases and distinguish from other. The magnetization or spin-spin correlation between the effective degrees of freedom is such a measure. In many occasion such correlation functions can only be calculated approximately. However the beauty of Kitaev model renders us to compute the correlation functions exactly \cite{smandal-prl}. Mathematically the correlation between two observables or operators $\hat{O}_1$ and $\hat{O}_2$ is expressed as $<\hat{O}_1 \hat{O}_2>$ where $<...>$ denotes ground state expectation value at zero temperature \cite{sakurai}. At finite temperature it means a thermal average. Physically it means what is joint probability that if the operator $\hat{O}_1$ takes value $O_1$ and the operator $\hat{O}_2$ takes value $O_2$. Apart from correlation function magnetization is also used to characterizes phases of magnetic and interacting spin systems. Magnetization is defined as the average value of magnetic moment in a system and at low temperature for quantum mechanical system it is defined as $< \vec{M} >$ where $\vec{M}= \frac{1}{N} \sum_i \vec{m}_i$ and  the angular bracket implies expectation with respect to ground state. Here $\vec{m}$ is the magnetic moment at a given site. Now we  follow an exact calculation of magnetization and two spin correlation function to find out  what kind of order is exhibited by the ground state wave function. To do that we first discuss an extension of Kitaev's Majorana fermionisation such that the mathematical steps to calculate the spin-spin correlation function or magnetization is very straightforward.  

 
\subsection{\label{bond-ferm}Bond fermion formalism}
   We have  seen in  Sec.~\ref{fso2.1} that  two complex fermions yield four Majorana fermions. Each complex fermion can be rewritten into two Majorana fermions. Now to facilitate the easy computation of spin-spin correlations we  invert the above procedure by regrouping two different Majorana fermions to define a complex fermion. We have seen that at every link there has been one conserve quantity named $u^{\alpha}_{i,j}$  made out of the Majorana fermion $c^{\alpha}_{i,a}$ and $ c^{\alpha}_{j,b}$. Here `$i$' and `$j$' denotes the two sites of a bond, `$a$' and `$a$' denotes sub-lattice indices and `$\alpha$' denotes a specific bond($\alpha=x,y,z$). We  regroup these two Majorana fermions to define a complex fermion named $ \chi_{\langle ij \rangle_{\alpha}}$ which lives on the bond joining sites `$i$' and `$j$'. We call this procedure as bond fermion formalism. From now on we follow the convention that the site `$i$' in the bond  $\langle ij\rangle_{\alpha}$  belongs to $a$ sub-lattice and the site `$j$'  belongs to $b$ sub-lattice. Also from now on we do  not mention the sub-lattice index `$a$' and `$b$' explicitly. We define complex fermions on each bond as,
\begin{eqnarray}
\label{chiadef1}
\chi_{\langle ij\rangle_{\alpha}}&=&\frac{1}{2}\left(c_i^{\alpha}+ic_j^{\alpha}\right)\\
\label{chiadef2}
\chi^\dagger_{\langle ij\rangle_{\alpha}}&=&\frac{1}{2}\left(c_i^{\alpha}-ic_j^{\alpha}\right)
\end{eqnarray}
\begin{center}
\begin{figure}[h!]
\center{\includegraphics[width=0.4\textwidth]{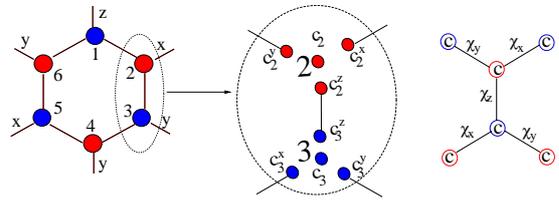}}
\caption{\label{bndfm}Elementary hexagon and `bond fermion' construction. A spin is replaced with four Majorana fermions $\left(c,c^{x},c^{y},c^{z}\right)$.  Bond fermion $\chi_{\langle23\rangle}$ for the bond joining site 2 and site 3 is shown . Spin operators are also defined.}
\end{figure}
\end{center}

For example with reference to the Fig.~\ref{bndfm}, for the $z$-bond joining site 2 and site 3, and for the $y$-bond joining site 1 and site 2, we define,

\begin{eqnarray}
\chi_{\langle23\rangle_z}&&=(c^{z}_2 +ic^{z}_3)  \\
\chi_{\langle12\rangle_y}&&=(c^{y}_1 +ic^{y}_2)
\end{eqnarray}
Then it follows that for the site `2' and `3' the $\sigma^{z}$ operator becomes,
\begin{eqnarray}
\sigma^{z}_2 &&= ic_2(\chi_{\langle23\rangle_z} + \chi^{\dagger}_{\langle23\rangle_z}) \\
\sigma^{z}_3 &&= c_2(\chi_{\langle23\rangle_z} - \chi^{\dagger}_{\langle23\rangle_z})
\end{eqnarray}

Below we write the  result of this re-fermionisation for a bond of type `$\alpha$' joining site `$i$' and `$j$',
\begin{eqnarray}
\label{chif}
&&\chi_{\langle ij\rangle_{\alpha}}=\frac{1}{2}\left(c_i^{\alpha}+ic_j^{\alpha}\right) \\
&&\sigma_i^{\alpha}=ic_i\left(\chi_{\langle ij\rangle_{\alpha}}
+\chi^\dagger_{\langle ij\rangle_{\alpha}}\right) \\
&&\sigma_j^{\alpha}= ic_j\left(\chi_{\langle ij\rangle_{\alpha}}
-\chi^\dagger_{\langle ij\rangle_{\alpha}}\right)
\end{eqnarray}
It is clear that three components of a spin operator at a given site gets connected to three  different $\chi$ fermions defined on the three different bonds emanating from it. The bond variables are related to the number operators of these fermions,
${\hat u}_{\langle ij\rangle_{\alpha}}\equiv ic_i^{\alpha}c_j^{\alpha} = 
2\chi^\dagger_{\langle ij\rangle_{\alpha}}\chi_{\langle ij\rangle_{\alpha}}-1$. Thus the effective picture is understood easily from the Fig.~\ref{bndfm}.  We identify  a $\chi$ fermion on every bond whose occupation number can be zero or one. This occupation number determines the value of $u_{\langle ij \rangle}$ on that bond. But these fermions are conserved and  serve as an effective $Z_2$ gauge potential for hopping `$c$' fermions. As $\chi$ fermions are conserved, all eigenstates can therefore be chosen to have a definite $\chi$ fermion occupation number. The Hamiltonian is then block diagonal in occupation number representation, each block corresponding to a distinct set of $\chi$ fermion occupation numbers for every bonds. Thus all eigenstates in the extended Hilbert  space take the following factorised form,
\begin{eqnarray}
\label{es1}
\vert{\tilde \Psi}\rangle =  \vert {\cal M}_{\cal G};{\cal G }\rangle & \equiv &
\vert {{\cal M}_{\cal G}}\rangle
\vert{\cal G }\rangle\\
\label{es2}
{\rm with}~~ \chi^\dagger_{\langle ij\rangle_{\alpha}}\chi_{\langle ij\rangle_{\alpha}}
\vert{\cal G} \rangle
&=&n_{\langle ij\rangle_{\alpha}}
\vert \cal G \rangle 
\end{eqnarray}
where $n_{\langle ij\rangle_{\alpha}}=(u_{\langle ij\rangle_{\alpha}}+1)/2$ and 
$\vert{\cal M}_{{\cal G}}\rangle$ is a many body eigenstate in the matter 
sector determined by `$c$' fermions, corresponding to a given $Z_2$ field configuration determined by $ \vert {\cal G} \rangle $. \\
\indent
\begin{figure}[h!]
\psfrag{a}[cb][cb]{$|\psi \rangle$}
\psfrag{c}[cb][cb]{$\sigma^{z}_{i} $}
\psfrag{b}[cb][cb]{$|\psi^{\prime} \rangle$}
\hspace{1.5cm}\includegraphics[width=0.8\linewidth]{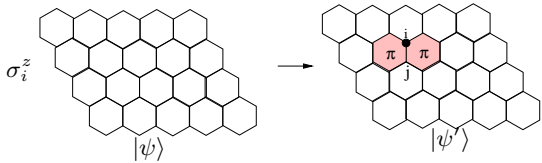}
\caption{\label{spflx}How a spin fractionalises into two static $\pi$ fluxes and a dynamic Majorana fermion is shown. $ |\psi \rangle$ is a state with zero flux. We apply $\sigma^{z}_{i} $ where site `$i$' is connected with site `$j$'. As a result we get a state $|\psi^{\prime} \rangle$ with two static $\pi$ fluxes at the plaquette sharing bond $\langle ij \rangle$ and a dynamic Majorana fermion represented by black circle.}
\end{figure}

\indent
Now we discuss the results of the above transformations and find that it brings an immediate visualization of the spin-operators. We observe  form Eq.~\ref{chif} that the effect of $\sigma^{\alpha}_i$ on any eigenstate becomes very clear. When the spin-operator acts on any eigenstate, in addition to adding a Majorana fermion at site `$i$', it changes  the bond fermion number from zero to one and vice versa  (equivalently, $u_{\langle ij \rangle_{\alpha}} \rightarrow - ~ u_{\langle ij \rangle_{\alpha}}$), at the bond $\langle ij \rangle_{\alpha}$. The end result is that one $\pi$ flux  is added to each of the two plaquettes that are shared by the bond 
$\langle ij \rangle a$ (Fig. ~\ref{spflx}). Mathematically the action of any spin-operator on any eigenstate can thus be written as, 
\begin{eqnarray}
\label{pinot}
\sigma_i^{\alpha} = ic_i\left(\chi_{\langle ij\rangle_{\alpha}}
+\chi^\dagger_{\langle ij\rangle_{\alpha}}\right) ~~\rightarrow ~~ ic_i~
{\hat\pi}_{1\langle ij \rangle_{\alpha}}~{\hat\pi}_{2\langle ij \rangle_{\alpha}}
\end{eqnarray}
with ${\hat \pi}_{1\langle ij \rangle_{\alpha}}$ and  ${\hat \pi}_{2\langle ij \rangle_{\alpha}}$ defined as operators that creates additional $\pi$ fluxes to two adjacent plaquettes shared by the bond $\langle ij \rangle_{\alpha}$ (Fig.~\ref{spflx}). Now it is easy to understand the action of one more spins which is connected with the previous ones. It yields ${\hat \pi}_{1\langle ij \rangle_{\alpha}}^2 = 1 $, since adding two $\pi$ fluxes is equivalent to adding (modulo $2\pi$) zero flux. This signify that while action of single spin operator creates the gauge fermion occupation number to change (either decrease or increase by one),  gauge fermion occupation number can be brought back to  initial values by action of a the same spin or a different  neighbouring spins. Only criteria is that the spin angular component of both the spins has to be same and this is determined by the nature of the bonds they are connected with. It is now straightforward to understand that two states with different flux configurations has vanishing overlap as they belong to different distribution of $\chi$ fermion occupation numbers. mathematically this implies that,
\begin{eqnarray}
\langle \mathcal{G} | \mathcal{G}^{\prime} \rangle = \delta_{n_{\mathcal{G}}, n_{\mathcal{G}^{\prime}}},
\label{nnpr}
\end{eqnarray}
where $ n_{\mathcal{G}}$ and $n_{\mathcal{G}^{\prime}} $ represent the distribution of $\chi$ fermions for the state $| \mathcal{G} \rangle$ and $| \mathcal{G}^{\prime} \rangle$ respectively. This observation will be extremely helpful to compute spin-spin correlations exactly. Not only two -spin correlation function, magnetization can also be calculated exactly. Apart from two spin correlation functions other multi spin correlations can be calculated with straight forward generalisations of this fact that for any  spin-spin correlator to be non-zero the first necessary conditions is that the simultaneous action of all the spin operators on the ground state must not change the flux configurations or equivalently must not alter the gauge fermion  occupation number of the bonds. This fact can be extended to all the eigenstates of the Kitaev model as well which is remarkable  for Kitaev model. \\
\subsection{Magnetisation}
\label{sec-mag}
Magnetization $\vec{\mathcal{M}}$ is an important physical quantity which is easily measurable and  can be controlled experimentally as well. A particular component of  magnetisation  is defined as  below,

\begin{eqnarray}
\label{mageq}
\mathcal{M}_{\alpha}&&= \frac{1}{\mathcal{N}}  \sum_i  \langle \sigma^{\alpha}_{i} \rangle  
\end{eqnarray}

\begin{figure}
\includegraphics[width=0.8\linewidth]{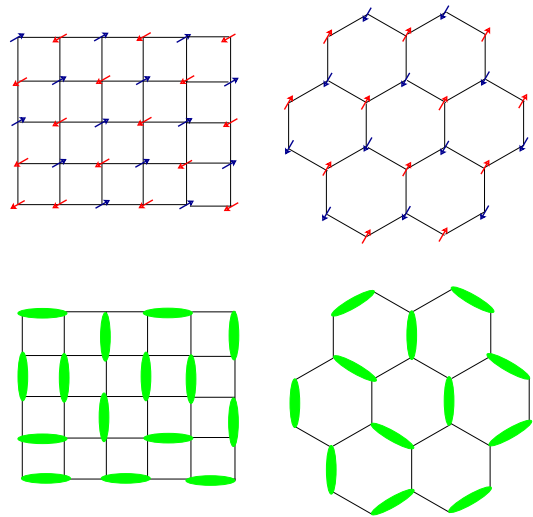}
\caption{\label{spliq}In the top panel we have shown anti-feromagnetic spin configurations in square lattice and in honeycomb lattice. In the below we have shown a dimer state configurations on square and honeycomb lattice where the green ellipses represents a dimer. The particular arrangement of dimers to cover the every site of the lattice is not unique and there exist alternative arrangement as well. The alternative arrangements often are topologically different.}
\end{figure}

In the above $ \langle \sigma^{\alpha}_{i} \rangle $ denotes the expectation value with respect to ground state. For ferromagnetic  state it can be found that at least one component of $\langle \sigma^{\alpha}_{i} \rangle  $ is non-zero at every site and it is identical for every site. On the other hand for anti-ferromagnetic state $\langle \sigma^{\alpha}_{i} \rangle$ are also non-zero at every site however their value is opposite at different sites and makes some pattern depending on the underlying structure. In Fig.~\ref{spliq} we have shown such ferromagnetic and antiferromagnetic structure in square and honeycomb lattice. We can see easily that for anti ferromagnetic state $ \mathcal{M}_{\alpha}$ is zero though for all the sites having
red spins are having opposite magnetization of blue spins. At a given site the average value of spin momentum is not zero. In the lower panel we have shown a different  state where at each shaded green region defined on a pair of sites has the following singlet state $|s \rangle = \frac{1}{\sqrt{2}} (| \uparrow \downarrow \rangle - |\downarrow \uparrow \rangle )$. For such a state average value of $\sigma^{\alpha}_i=0$ at any site \cite{moessner,philip,alain}. This state is fundamentally different than the anti ferromagnetic state.
For the antiferromagentic state at a given site $\sigma^{\alpha}_i$ is not zero. Now let us see what is the value of $\langle \sigma^{\alpha}_i \rangle =0 $ for the  ground state of the Kitaev model. From the definition of spin-operator as expressed in Eq.~\ref{pinot} we see that action of a spin on the ground state creates two additional flux in the ground state as shown in Fig.~\ref{spflx}, in addition it also adds a Majorana fermion to the ground state. Mathematically this is expressed as 
\begin{equation}
\label{mageq2}
\sigma_i^{\alpha}\vert{\cal G}\rangle\vert{\cal M}_{\cal G}\rangle=
c_i\vert{\cal G}^{i\alpha}\rangle\vert{\cal M}_{\cal G}\rangle
\end{equation}

Now as the different flux configuration state is mutually orthogonal due to  different occupation number  of the conserved $\chi$ fermion, we obtain,

\begin{eqnarray}
&&\langle {\cal G} \vert \langle {\cal M}_{\cal G} \vert \sigma_i^{\alpha} \vert{\cal G}\rangle \vert{\cal M}_{\cal G}\rangle = \langle {\cal G} \vert \langle {\cal M}_{\cal G} \vert  c_i \vert{\cal G}^{i \alpha}\rangle \vert{\cal M}_{\cal G}\rangle =0 
\end{eqnarray}

because $ \langle {\cal G} \vert{\cal G}^{i\alpha}\rangle =0 $ following Eq.~\ref{nnpr}. Thus we see that the magnetization is zero for Kitaev model. It is zero for every site unlike the AFM state where the magnetization at a given site is not zero but when averaged over the system it is zero. For spin singlet also the magnetization is zero at a given site.   However there is an important difference between the AFM state, and the spin-singlet state shown in Fig.~\ref{spliq}. For AFM state $\langle \sigma^z_{i} \sigma^z_j \rangle = \pm 1$ depending on weather the site `$i$' and `$j$' are both  are aligned in the same direction or in opposite direction. Note that it does not depends on the distance between the two sites. It is said that the state has a long range ordered state.  For singlet state $\langle \sigma^z_i \sigma^z_j \rangle$ is not zero if the site `$i$' and `$j$' both belongs to same singlet otherwise it is zero.  Thus there is no long range correlation in the singlet product state.  However for the singlet state $\langle \sigma^{\alpha} \rangle$ is zero which  bears similarity with Kitaev model. However there are important differences between the singlet state and Kitaev model which will be established once we calculate the two spin correlation function. With the knowledge of few magnetic states as discussed above, we now move on to calculate the spin-spin correlation function of Kitaev model which  would establish its spin-liquid nature. 
\begin{center}
\Large{Exercise-7}
\end{center}

{\it Consider a spin-silglet state  formed between site 1 and site 2 as expressed by $|s \rangle= \frac{1}{2} \left( |\uparrow_1 \rangle  |\downarrow_2 \rangle -   |\downarrow_1 \rangle  |\uparrow_2 \rangle \right)$ where $| \uparrow_i \rangle$ and $| \downarrow_i \rangle$ refers the spin at a site `$i$' oriented along positive and negative $z$ axis respectively. Do the following questions.} 
\begin{itemize}
\item{\it { Calculate $ \langle s | \sigma^z_1 \sigma^z_2 | s \rangle$}}
\item{\it{ Using the transformation $ | \uparrow_i \rangle = \frac{1}{\sqrt{2}} \left( | \rightarrow_i \rangle + | \leftarrow_i \rangle \right), {\rm and} | \downarrow_i \rangle = \frac{1}{\sqrt{2}} \left( | \rightarrow_i \rangle - | \leftarrow_i \rangle \right)$ where $| \rightarrow_i \rangle {\rm and} | \leftarrow_i \rangle $ denotes the spin-oriented along $x$-axis, express $| s \rangle$ in terms of $| \rightarrow_i \rangle {\rm and} | \leftarrow_i \rangle $ with $i=1,2$.}}
\item{\it{ Now calculate $\langle s| \sigma^x_1 \sigma^x_2| s \rangle $ and show that it is non-zero. Similarly find $\langle s| \sigma^y_1 \sigma^y_2| s \rangle $ and comment on the symmetry property of the state $|s \rangle$}}
\end{itemize}
From the above exersize it must be clear that spin-singlet product state is an unique state with no long range correlation but short range nearest neighbour correlation exist with $\langle \sigma^{\alpha}_i \sigma^{\alpha}_j  \rangle$ non-zero if `$i$'
and `$j$' are nearest neighbour and they belong to a given dimer.

\subsection{Two-spin correlations, fractionalization, de-confinement}

Here we give an outline of computation of  spin-spin correlation functions in Kitaev model which is valid for any region of the phase diagram. The derivation here is given in the extended Hilbert space but can be easily extended without any difficulties to physical Hilbert space. This happens because after the implementation of projection operator to ground state  wave function obtained in extended Hilbert space  one obtains many daughter states differed with respect each other only in  gauge sector by fermion occupation number in certain manner so that action of two spins on any of them still creates orthogonal states and mutual overlap between such states are again vanishes.  The above fact is expressed technically as the following. Because the spin operators are gauge invariant (commute with the projection operator) the  result of computation of spin-spin correlation function in the extended Hilbert space yields no problem at all.  First to begin with, we consider the two spin dynamical correlation  functions, in an arbitrary eigenstate of the Kitaev Hamiltonian in some fixed gauge field configuration ${\cal G}$,
\begin{equation}
\label{d2scf1}
S_{ij}^{\alpha \beta}(t)= \langle{\cal M}_{\cal G}\vert\langle{\cal G}\vert 
\sigma_i^{\alpha}(t)\sigma_j^{\beta (0)}
\vert{\cal G}\rangle\vert{\cal M}_{\cal G}\rangle
\end{equation}
Here $A(t) \equiv e^{iHt} A e^{-iHt}$ is the Heisenberg representation of an 
operator A, keeping $\hbar = 1$.  Physically the  quantity $ S_{ij}^{\alpha \beta}(t)$ in Eq.~\ref{d2scf1} gives the  joint probability amplitude  of finding a spin at `$j$' (at time zero) along  `$\alpha$' axis and of finding another spin at `$i$' to be  along `$\beta$' axis.  As discussed before we write the action of spin operator on any eigenstate as, 
\begin{eqnarray}
\label{sigones1}
\sigma_j^{\beta(0)}\vert{\cal G}\rangle\vert{\cal M}_{\cal G}\rangle&=&
c_j(0)\vert{\cal G}^{i\beta}\rangle\vert{\cal M}_{\cal G}\rangle\\
\label{sigones2}
\sigma_i^{\alpha (t)}\vert{\cal G}\rangle\vert{\cal M}_{\cal G}\rangle&=&
e^{i(H-E)t}c_j(0)\vert{\cal G}^{i\alpha}\rangle\vert{\cal M}_{\cal G}\rangle
\end{eqnarray}
where, $\vert{\cal G}^{i\alpha (j\beta)}\rangle$ denote the states with extra
$\pi$ fluxes added to ${\cal G}$ on the two plaquette adjoining the 
bond in $\alpha$ or $\beta$  directions. It means that if $\alpha=x$, then the 
two plaquettes are created adjoining the site `$i$' to another  site `$k$' joined
by $x$-bonds. Similar explanation goes for the $\beta$ indices as well. In the above equation
$E$ is the energy eigenvalue of the eigenstate 
$\vert{\cal G}\rangle\vert{\cal M}_{\cal G}\rangle$. Since the $Z_2$ fluxes on 
each plaquette is a conserved quantity due to the fact the it is determined by the occupation number
of gauge fermions ($\chi$ fermions) on the links, it is obvious that the correlation
function in Eq.~\ref{d2scf1} which is the overlap of the two 
states in equations \ref{sigones1}, \ref{sigones2} is zero unless 
the spins are on neighbouring sites indicating that for non-vanishing values of correlation function  we must
have `$j$' as the nearest neighbour site of the site `$i$'. Also note that the components $\alpha$ and $\beta$ must be identical so
that the effect of $\sigma^{\alpha}_i$ and $\sigma^{\beta}_j$ is to  create a pair of flux and then annihilate it bringing the flux configuration same as before. The above simple observations says that the
dynamical spin-spin correlation has the form,
\begin{eqnarray}
\label{d2scfres1}
S_{ij}^{\alpha \beta}(t)&&=g_{\langle ij \rangle_{\alpha}}(t)\delta_{\alpha,\beta},{\rm ~for}~i,j {\rm~~nearest~neighbours}~~~~~~~ \\
&&= 0 ~~~~~~~~~~~~~~~~~~~{\rm otherwise.}
\end{eqnarray}
Computation of $g_{ij}(0)$ is  straightforward 
in any eigenstate $|\cal M_{\cal G}\rangle$. For the ground state 
where conserved $Z_2$ charges are unity at all plaquette, we have an analytical expressions of the ground state wave functions as given in Eq.~\ref{gr-st}.  For simplicity  we provide the expression for equal-time correlation by having $t=0$ without loss of generality. For correlation function at different time interested readers are requested to consult the original article \cite{smandal-prl}. Using Eq.~\ref{sigones1} and Eq.~\ref{sigones2}, the equal time 2-spin correlation function for the 
bond $\langle ij \rangle_{\alpha}$ reduces to the following simple expectation value :

\begin{eqnarray}
\langle \sigma^{\alpha}_{i}\sigma^{\alpha}_{j}\rangle \equiv
S^{\alpha\alpha}_{\langle ij \rangle_{\alpha}}(0)
= \langle{\cal M}_{\cal G}\vert\langle{\cal G}\vert 
c_i c_j
\vert{\cal G}\rangle\vert{\cal M}_{\cal G}\rangle
\label{cicj}
\end{eqnarray}

In the above equation we have omitted the time index for $c_i$ and $c_j$ operator for simplicity. One can simplify the expressions in Eq.~\ref{cicj} by using the Fourier transformations of $c_i$ and $c_j$ operators and then employing the orthogonal transformations for $c_k$  operators as given in Eq.~\ref{kham1}, we obtain the following final expressions for the two-spin equal time correlation function,
\begin{eqnarray}
\nonumber
\langle \sigma^{\alpha}_{i}\sigma^{\alpha}_{j}\rangle \equiv
S^{\alpha\alpha}_{\langle ij \rangle_{\alpha}}(0)
=\frac{\sqrt{3}}{16\pi^{2}}\int_{BZ} \cos\theta(k_{1},k_{2})dk_{1}dk_{2}
\end{eqnarray}
Where $\cos\theta(k_{1},k_{2})=\frac{\epsilon_{k}}{E_{k}}$, 
$ E_{k}=\sqrt{(\epsilon_{k}^{2}+\Delta^{2}_{k})}$, in the Brillouin zone. 
$\epsilon_{k}=2(J_{x} \cos k_{1} +J_{y}\cos k_{2} +J_{z})$,  
$\Delta_{k}=2(J_{x}\sin k_{1}+J_{y}\sin k_{2})$,
$k_{1}=\textbf{k}.\textbf{n}_1$, $k_{2}=\textbf{k}.\textbf{n}_2$ and 
$\textbf{n}_{1,2}=\frac{1}{2}\textbf{e}_{x} \pm
\frac{\sqrt{3}}{2}{\textbf{e}_{y}}$ are unit vectors along $x$ and $y$ bonds.
At the point, $J_x=J_y=J_z$, we get 
$S^{\alpha\alpha}_{\langle ij \rangle_{\alpha}}(0) = -0.52 $.
\begin{figure}
\includegraphics[width=0.6\linewidth]{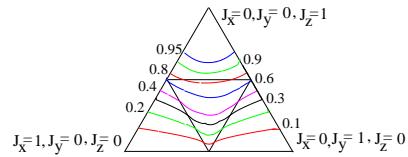}
\caption{\label{zzcore0}Contour plot of nearest-neighbour $z-z$ correlation in the parameter space. We note that $z-z$ correlation increases as magnitude of $J_z$ increases.}
\end{figure}
It is of interest to  examine how this non-vanishing two spin correlation depends on the
parameters of the Hamiltonian mainly on $J_x, J_y, J_z$ The Fig.~\ref{zzcore0} shows 
how the nearest neighbour $z-z$ correlations varies in the parameter ( $J_x , J_y, J_z$ ) space.
As we have calculated the above two spin-correlation function for two spins connected by a $z$-bond,
we observe that as we increase the magnitude of $J_z$, the value of correlation also increases.
In particular two limiting cases in the parameter values is worth examining. Consider the case
$J_z=0$. Looking at the Fig.~\ref{fig1} we observe that on this limit, the honeycomb lattice
reduces as collection of decoupled chains. Thus  the two spins which  have been joined by an interaction
are now  without any interaction between them, though the spins maintain their original interactions with
nearest neighbour spins in the respective chains. As there is no interaction between these two spins, we
expect that there will be no correlation between them and from Fig.~\ref{zzcore1}, we do find that
indeed the correlation between them is zero. The other limiting cases is $J_z= \infty$ or $J_x=J_y=0$, in this
limiting case the two spins connected by a `$z$' bonds does not interact with any other spins except the one
connected by $J_z$ bonds. Thus one expect that the correlation will be stronger and reach the saturation value one
as shown in Fig.~\ref{zzcore1}. Another interesting fact to note that there is little change in slope as we increase 
from gapped region to gapless region. \\
\begin{figure}
\psfrag{d}{0.2}
\psfrag{f}{$J_z$}
\psfrag{b}{1}
\psfrag{c}{0.8}
\psfrag{a}{$0,0$}
\psfrag{e}{$\langle \sigma^z_i \sigma^z_j \rangle$}
\label{zzcore}
\includegraphics[width=0.6\linewidth]{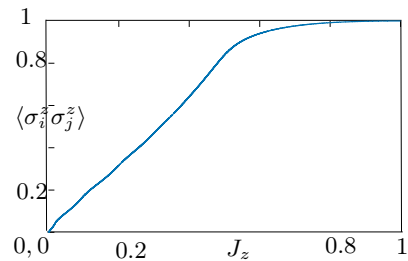}
\caption{\label{zzcore1} In the above we plot the $\langle \sigma_{1z} \sigma_{2z} \rangle $ on a give $z$-bonds.
We have taken $J_x=J_y=0.5$ and vary $J_z$ from 0 to 1. As expected for $J_z=0$ correlation is zero and for large values it saturates to 1. }
\end{figure}
\\
\indent 
\vspace{0.5cm}
\\
\textit{\bf{Fractionalization and De-confinement}:}  We note that here we have only discussed the derivation of equal time correlation function which means that in Eq.~\ref{d2scf1} both $\sigma$'s have identical argument  for time. We found that the spin-spin correlation function is short-range and also bond dependent. This property of short-range and bond dependent nature of correlation function remains unchanged for different time correlation function as well. However the derivation and the exact expression  requires a little mathematical digression and for that we do not present that in this article. Instead  we discuss the issue without going into technical details to meet the inquisitiveness of the reader. Let us try to understand  the physical consequences of Eq.~\ref{d2scf1}, Eq.~\ref{sigones1} and ,Eq.~\ref{sigones2}. The r.h.s of Eq.~\ref{d2scf1} involves inner product of two parts which are $\sigma_j^{\beta (0)}\vert{\cal G}\rangle\vert{\cal M}_{\cal G}\rangle$ and  $ \langle{\cal M}_{\cal G}\vert\langle{\cal G}\vert \sigma_i^{\alpha}(t)$ which are shown in Eq.~\ref{sigones1} and Eq.~\ref{sigones2}. The meaning of Eq.~\ref{sigones1} is easy to understand which tells what happens  when a given spin acts on an eigenstate. It  creates two fluxes in the gauge sector  $|\mathcal{G} \rangle$ of the eigenstate and also adds a Majorana fermion in the matter sector( $| \mathcal{M}_{\cal G} \rangle$). Let us try to expand the meaning of Eq.~\ref{sigones2} which represents how the action of a given spin on an eigenfunction evolves as function of time. To this end we yield the intermediate steps reaching Eq.~\ref{sigones2} as follows.

\begin{eqnarray}
&& \sigma_i^{\alpha}(t) \vert{\cal G}\rangle\vert{\cal M}_{\cal G}\rangle \nonumber \\
&&= e^{i H t} \sigma_i^{\alpha}(0) e^{-iHt} \vert{\cal G}\rangle\vert{\cal M}_{\cal G} \rangle \nonumber \\
&&= e^{i H t} \sigma_i^{\alpha}(0) e^{-iE t} \vert{\cal G}\rangle\vert{\cal M}_{\cal G} \rangle \nonumber \\
&&= e^{i (H-E)t } \sigma_i^{\alpha}(0) \vert{\cal G}\rangle\vert{\cal M}_{\cal G} \rangle \nonumber \\
&&= e^{i (H-E)t } c_i(0) \vert{\cal G}^{i,\alpha} \rangle \vert{\cal M}_{\cal G} \rangle .
\end{eqnarray}

\begin{center}
\begin{figure}[h!]
\psfrag{a}[cc][cc]{$A_{ij}(t)$}
\psfrag{k}[cc][cc]{$k$}
\psfrag{j}[cc][cc]{$j$}
\psfrag{j}[cc][cc]{$j$}
\psfrag{m}[cl][cl]{Majorana Fermion}
\center{\hbox{\epsfig{figure=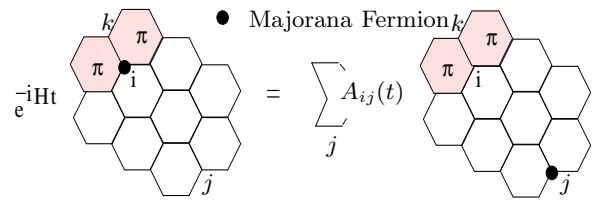,height=1in,width=3in}}}
\caption{\label{fractmj1}Time evolution and fractionisation of a spin flip at t = 0 at 
site `$i$', into a  $\pi$-flux pair and a propagating Majorana fermion.}
\end{figure}
\end{center}

In the first and second steps we have used the  definition of time evolution of an operator and applied eigenstate condition with energy $E$ respectively. The third step is a simple rearrangement  and fourth step implements the effect of spin $\sigma^{\alpha}_i$ which is equivalent to create two fluxes in the $\vert{\cal G}\rangle $ (to have $\vert{\cal G}^{i,\alpha} \rangle $) and also adding a Majorana fermion $c_i$ in the matter sector $\vert{\cal M}_{\cal G}$. Now We need to consider the effect of the exponential operator $ e^{i (H-E)t }$ on this. Remember that from Eq.~\ref{mageq2} that $c_i(0) \vert{\cal G}^{i,\alpha} \rangle \vert{\cal M}_{\cal G} \rangle $ is not an eigenstate of $H$. The  situation is complicated  for two reason. Firstly $ \vert{\cal G}^{i,\alpha} \rangle $  makes all the conserved quantity '$u$' to be 1 except on a bond connecting site `$i$'. Secondly $ c_i(0) \rangle \vert{\cal M}_{\cal G} \rangle $ is not an eigenstate for such distribution of $u$'s in Eq.~\ref{3}. In general if a state is not an eigenstate of an Hamiltonian then the action of Hamiltonian on it results into superposition of many other eigenstates. Thus the action $ \sigma_i^{\alpha}(t) \vert{\cal G}\rangle\vert{\cal M}_{\cal G}\rangle$ can be mathematically represented as follows,
\begin{eqnarray}
&& \sigma_i^{\alpha}(t) \vert{\cal G}\rangle\vert{\cal M}_{\cal G}\rangle \nonumber \\
&&=  e^{-iE t} \left( \sum c_j(t) \mathcal{A}_{ij} \vert{\cal M}_{\cal G}\rangle_j \right) \vert{\cal G}^{i,\alpha} \rangle,  
\end{eqnarray}
where $ \vert{\cal M}_{\cal G}\rangle_j$ represents `$j$'the eigenstate in the presence of two fluxes. In each of these $\vert{\cal M}_{\cal G}\rangle_j$, the fluxes are in identical position, they are static but the Majorana fermion added
to each of this eigenstate $\vert{\cal M}_{\cal G}\rangle_j$ are in different position in general. This effect is a
remarkable effect and  signifies that in essence a given spin is composed of two objects, a Mojorana fermion and a two flux. While the fluxes  do not move, the Majorana fermion moves as time evolves and it is not confined in a given region unlike  static flux pairs. This implies that physically a spin is fractionalized into two objects. The fact that Majorana fermion is moving gives rise to a de-confinement phenomena. In Fig.~\ref{fractmj1}, we depict pictorially the phenomena of fractionalization and de-confinement. The exact formula for $\sigma_i^{\alpha}(t) \vert{\cal G}\rangle\vert{\cal M}_{\cal G}\rangle $ can be seen in the original manuscript \cite{smandal-prl}.
 
Now in comparison with the singlet state we discussed before we find a number of differences. The similarity between the two is that in both the cases the two spin correlation function is short range i.e it exists only for nearest neighbour bonds only. However there are a number of important differences. For example for singlet state, the correlation function is non-zero if the two sites belong to a singlet only and for  a given singlet $\langle \sigma^{\alpha}_{i} \sigma^{\alpha}_j \rangle$ exist for $\alpha=x,y,z$. However for Kitaev model for any two nearest neighbour correlation function is non-zero but they are anisotropic in the sense that for a `$\alpha$' type bonds only  $\langle \sigma^{\alpha}_{i} \sigma^{\alpha}_j \rangle$ is non-zero and for other component the spin-spin correlation function is zero.

\subsection{Multispin Correlation function,topological degeneracy}
\label{multi}
\begin{figure}
\includegraphics[width=0.6\linewidth]{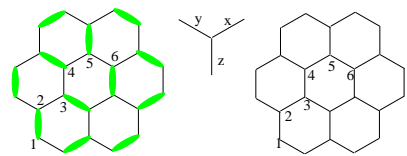}
\caption{\label{mulsp}Multispin correlation function exist for Kitaev model. For the sites 1 to 6, $\langle \sigma^z_1 \sigma^y_2 \sigma^y_3 \sigma^y_4 \sigma^z_5 \sigma^y_6 \rangle$ is non-zero. For detail see text below. }
\end{figure}
We have seen that for Kitaev model magnetization is zero at every site. Also two spin correlation function is extremely short range such that it does not extend beyond nearest neighbour. It is natural to ask then what is the difference between a normal paramagnetic or other disordered magnetic state  for which long range two spin correlation is zero. We will now  discuss that. Though the two spin correlation exists only for nearest neighbour spins there is underlying long range correlation between spins such that spins at long distances are entangled unlike paramagnetic or other disordered  magnetic state where such long range correlation does not exist. We will also compare it with the singlet state described before and find a very important differences. The existence of long range multispin correlation function depends again on  the creation and annihilation of flux configurations in  $| \mathcal{G} \rangle$ such that the final and the initial flux configurations remains same. We note that this was the main  reason for non-vanishing two spin correlations. In the right panel of Fig.~\ref{mulsp}, we have drawn a cartoon of Kitaev model where $x,y$ and $z$ type of bonds are shown. Various numerics 1 to 6 denote many spins for which we wish to show that a non-vanishing multispin correlation function exist.  We notice that for bond $1-2$, $\sigma^z_1 \sigma^z_2$   does not change the flux configurations in $| \mathcal{G} \rangle$, similarly   or bond $2-3$, $\sigma^x_2 \sigma^x_3$ does not change the flux configuration in $| \mathcal{G} \rangle$. Thus for the site $1-2-3$, $\sigma^z_1 \sigma^z_2 \sigma^x_2 \sigma^x_3 \sim \sigma^z_1 \sigma^y_2 \sigma^x_3$ does not change the flux configurations in $| \mathcal{G} \rangle $ and this three spin correlator is non-zero. Thus for the spin 1 to 6 the following multispin correlation is non-zero,
\begin{equation}
S_{1-6}= \sigma^z_1 \sigma^y_2 \sigma^y_3 \sigma^y_4 \sigma^z_5 \sigma^y_6
\end{equation}
\begin{figure}
\includegraphics[width=0.6\linewidth]{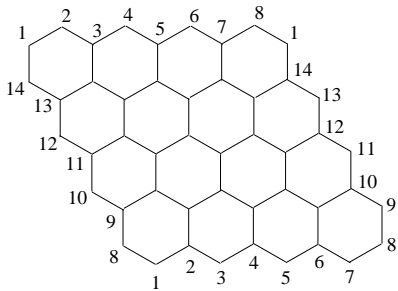}
\caption{\label{torus}A cartoon picture of torus geometry for a $4 \times 4$ hexagon has drawn. The numerics show the specific way the boundaries are connected with each other. The above geometry is not unique and  different geometry of lattice exist for torus realization. }
\end{figure}
Thus for any path of arbitrary length there is  a non-vanishing multispin correlation associated with it. It shows that all the spins irrespective of the distance between them are really entangled or depends on each other in a specific way. 
This is not true for ordinary paramagnetic or disordered spin configurations. Also the above multispin correlation is not present in the singlet state as shown in the left panel because there is no correlation exist between spin $2$
and spin $3$. However there could be an alternative singlet covering such that singlet state is formed on bonds $2-3$, $4-5$ etc. If we consider  the superposition of this two states, we might have a multispin correlation non-zero. Thus though Kitaev model  and the spin-singlet state both has short range two spin-correlation of different nature as well as of long range multispin correlation of different nature. The long range multispin correlations or string operators lead to another novel concept which is
knew as topological order. The word topology here is used to denote the fact that the movement of a given spin really depends on each other and every spin is entangled every other spin in a particular manner. This also says that a spin can not be arbitraily oriented and it must conform to  certain global pattern or the spin-orientations of other spins. Such global pattern of spin configurations of all the spin might lead some global constraint such that certain patterns are mutually exclusive leading to an existence of long range  string order parameter which takes distinct values for each different pattern of spin configuration. We will now explain this fact for Kitaev model defined on torus of honeycomb lattice. Torus means two dimensional lattice such that it is periodically connected in both the direction. In Fig.~\ref{torus}, we have shown such periodic boundary condition applied to a $4 \times 4$ arrangement of hexagons. As can be seen that the lower and upper  sides are connected as well as the left and the right sides. Note  that each one dimensional edges are periodically connected. Now we  evaluate some numbers which the reader are requested to  cross-check of this $4 \times 4$ arrangement of hexagon having $32$ sites.  For a general $N \times N$ hexagons, there
are total number of sites $M= 2 N^2$. Total number of hexagon is also $M$. We know that associated with each hexagon there is a conserve quantity called $B_p$. Thus there are in total $M$ number of $B_p$. Now one can check that for torus geometry  once all the $B_p$
are  multiplied together it becomes unity such that,
\begin{eqnarray}
\prod_{p=1,M} B_p=1
\end{eqnarray} 
This tells that all the $B_p$ is not independent and there are in total $M-1$ independent $B_p$. It can be shown that there is a conserve quantity associated with every closed loop, but they are not independent as they can be constructed by a product of certain $B_p$'s. However there two global loop operator  which extends from one end to another which can not be expressed as a product of $B_p$ and there are two such loop operators which we call $W_1$ and $W_2$. These are analogous to Wilson loop operator in quantum field theory \cite{giles,wilson}. For example take a horizontal zig-zag line constituted out of alternating $x$ and $y$ bonds. the product of $\sigma_z$ for all the sites on that line is a conserved quantity.
\begin{eqnarray}
\label{wone}
W_1 = \prod_{x-y~chain} \sigma^z_i
\end{eqnarray}
Similarly for a $x-z$ chain extending from one end to another end, we can construct $W_2$ which is conserved. Thus total number of conserved quantity becomes $M+1$. Now it can be checked that $W^2_i=1$ implying that eigenvalues of $W_i$ is $\pm 1$. The presence of these two additional independent conserved quantities imply that there will be four different eigenstates of certain identical configurations of $B_p$ but with different eigenvalues of $W_i$. These four eigenfunctions can be represented as,
\begin{eqnarray}
&&|\Psi_1 \rangle= |\mathcal{M}_{\mathcal{G}} \rangle | \mathcal{G}(B_p); W_1=1, W_2=1 \rangle \\
&&|\Psi_2 \rangle= |\mathcal{M}_{\mathcal{G}} \rangle | \mathcal{G}(B_p); W_1=-1, W_2=1 \rangle \\
&&|\Psi_3 \rangle= |\mathcal{M}_{\mathcal{G}} \rangle | \mathcal{G}(B_p); W_1=1, W_2=-1 \rangle \\
&&|\Psi_4 \rangle= |\mathcal{M}_{\mathcal{G}} \rangle | \mathcal{G}(B_p); W_1=-1, W_2=-1 \rangle 
\end{eqnarray}
\begin{figure}
\includegraphics[width=0.8\linewidth]{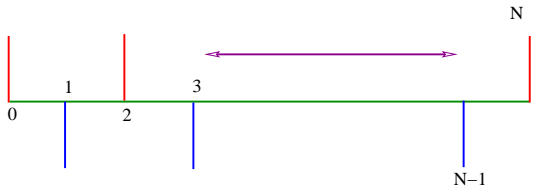}
\caption{\label{topoja} Various red lines represent the spin states with even number of  spin in the $+ z$ direction. The blue lines show odd number of spins in the $- z$ direction. The red states contribute to $W_1=1$ and the blue states contribute to $W_1= -1$.}
\end{figure}
  These $W_i$  manifests a hidden topological order which we will try to understand by considering how many ways $W_1$ defined in Eq.~\ref{wone} can be made one for one dimensional arrangement of spins. In Fig.~\ref{topoja}, we have shown by red line the states for which even number of spins are aligned in postive z-direction.  For a one dimensional chain of $N$ sites there are $2^{N-1}$ such states.  Similarly by blue line, we represents states with odd number of spins oriented along negative $z$-axis. All the red states in a given $x-y$ line contribute to the  wave function to make $W_1=1$ and such states can not be converted to an eigenfunction with $W_1=-1$ by local flipping of a given spin or a number of spins. For that we need to convert all the red states to the blue one. It is non-local transformation in the Hilbert space and thus topological. However the energy does not depends on the eigenvalues of $W_i$, it only depends on the flux configurations $B_p$. Thus all the states $| \Psi_i \rangle, i=1,4$ has identical energy in the thermodynamic limit \cite{saptarshi-2013}. This is a manifestation of topological degeneracy.

\section{Discussion}
\label{dis}
In this pedagogical paper, we have tried to explain few aspects of Kitaev model \cite{kitaev-2006}.  We have explained in detail how the Majorana fermionisation  is used to solve the model. By solving we  mean finding the eigenvalue and eigenvectors. All the eigenvectors and eigenvalues of the Kitaev model can be found in principle. This is a distinct feature which makes it stand apart from other  exactly solvable model where in some cases we have only the ground state \cite{majumder-ghosh}. It may be noted that Kitaev model can be defined in any lattices with co-ordination number of three and exact solution can be obtained \cite{saptarshi-2009,saptarshi-2014}. The Kitaev model is reduced to a $Z_2$ gauge theory and in the ground state sector  all the $Z_2$ gauge fields are such that their product over a plaquette becomes unity. Uniform choice of $Z_2$ gauge fields for every link is convenient choice for such configuration. The spectrum, phase diagram and calculation of correlation function  has been shown explicitly and elaborately. Notable feature of the phase diagram is that it consists of  gapless and gapped phase both. In the gapless phase near the region where the spectrum becomes zero, the energy varies linearly with momentum. On the hand where the spectrum is gapped, the energy varies as quadratically with momentum.  The correlation function calculation shows that  two spin correlation function is  short range and bond dependent. This  property of correlation function is  true for both static and dynamic correlation function. The short range nature of correlation function indicates that Kitaev model is an example of quantum spin liquid state where there is no long range order between two spins as well there is no average magnetization i.e average value of any spin operator is zero.    We have shown that the action of a spin operator on any eigenstate is two fold. Firstly it changes the occupation number of a $Z_2$ gauge field on  a particular bond by changing the value of $B_p$ in the adjacent plaquette. Secondly at also adds a Majorana fermion at a given site the spin belongs to. However the time evolution of the state shows that the pair of flux created does not move but the Majorana fermion added can move and is not confined or attached to that site only. This phenomena is called  fractionalization of spins into static fluxes and dynamic Majorana fermion. The moving Majorana fermion defines a de-confinement phase which is gapless one with linear  spectrum at low energy. \\

\indent
Moving further we have shown how the long range or multispin entanglement exist in an eigenfunction by showing  non-zero  multispin correlation function and explained how it is different than other magnetic state with short range two-spin correlation. Existence of topological degeneracy of eigenfunction in thermodynamic limit has been  explained schematically.  Historically Kitaev model was conceived  mainly with the aim of implementation to quantum computations due to the existence of topological order and non-trivial topological excitations called anyons. It may be noted that the gapped phase of Kitaev model realizes abelian anions and  gapless phase realizes non-abelian anyons. The braiding properties of these anyons in relation to realizing quantum gates has been discussed in detail \cite{kitaev-2003, kitaev-2006}. Kitaev model is exactly solvable in the sense that all the eigenfunction and eigenvalues can be obtained formally. The exact solvability is lost once other more conventional spin-spin interaction such as Ising or Heisenberg interaction \cite{saptarshi-2011,subhro-1,subhro-2} is added to Kitaev model. In this case the $Z_2$ gauge field defined on each bond  is no longer remains conserved. The  flux operator $B_p$ also does not commute with the Hamiltonian any more. Physically this means that the fractionalisation of spins into Majorana fermion and static fluxes  are not possible no longer. Because the fluxes also move along with the Majorana fermions though their dynamics may be at different scale. Though the Kitaev model was proposed with a theoretical foundation for quantum computations, it was rather surprising that this unusual interactions happens to exist in certain materials such as Iridiate and $RuCl_3$ etc \cite{Jac,jackeli,a-banerjee}. Unfortunately along with Kitaev interaction, there exists some other interactions such as Heisenberg type interactions. Presently intense research is underway to explore various aspects Kitaev model such as fractionalisation, de-confinement to confinement transition, detection of spin-liquid state  etc. Interested reader are requested to go through some of the recent developments \cite{takagi} and the references there in. \\
\indent
{ \bf Acknowledgement :} A. M. Jayannavar acknowledges DST, India for JCBose  fellowship. S M acknowledges G. Baskaran and R. Shankar for their  encouraging presence throughout the journey involving Kitaev model.

\end{document}